\documentclass[10pt,twocolumn,twoside]{IEEEtran}

\usepackage{graphicx}
\usepackage{cite}
\usepackage[cmex10]{amsmath}
\usepackage{amssymb}

\usepackage{tikz}
\usepackage{pgfplots}
\pgfplotsset{compat=1.3}
\usetikzlibrary{arrows,snakes,shapes}

\DeclareMathAlphabet{\mathbit}{OML}{cmr}{bx}{it}

\DeclareMathOperator{\E}{E}
\DeclareMathOperator{\T}{T}
\DeclareMathOperator{\Tr}{tr}

\DeclareMathOperator{\Probability}{Pr}
\DeclareMathOperator{\Diag}{diag}

\DeclareMathOperator{\Sinc}{sinc}

\renewcommand\vec[1]{\operatorname{vec}\left(#1\right)}
\renewcommand\arcsin[1]{\operatorname{arcsin}\left(#1\right)}

\DeclareMathOperator{\fieldR}{\mathbb{R}}

\DeclareMathOperator{\fieldN}{\mathbb{N}}

\DeclareMathOperator{\distN}{\mathcal{N}}

\newcommand{\diag}[1]{\Diag{\left(#1\right)}}

\newcommand{\sign}[1]{\Sign{\left(#1\right)}}

\newcommand{\ve}[1]{\boldsymbol{#1}}

\newcommand{\exdi}[2]{\E_{#1} \left[#2\right]}

\renewcommand{\exp}[1]{\operatorname{exp}\left(#1\right)}

\newcommand{\tr}[1]{\Tr \left(#1\right)}
\newcommand{\Prob}[1]{\Probability\left\{#1\right\}}

\newcommand{\sincb}[1]{\Sinc \left(#1\right)}

\newcommand\Sign{\operatorname{sign}}

\newcommand{\floor}[1]{\lfloor #1 \rfloor}

\title{Sensitivity Analysis for Binary Sampling Systems via Quantitative Fisher Information Lower Bounds}

\author{Manuel~S.~Stein
\thanks{This work was in part supported by the German Academic Exchange Service (DAAD) with funds from the German Federal Ministry of Education and Research (BMBF) and the People Program (Marie Sk{\l}odowska-Curie Actions) of the European Union's Seventh Framework Program (FP7) under REA grant agreement no. 605728 (P.R.I.M.E. - Postdoctoral Researchers International Mobility Experience). This work was also in part funded by the Deutsche Forschungsgemeinschaft (DFG, German Research Foundation) - grant no. 413008418 (Research Fellowship Program).
}
\thanks{M. S. Stein is with the Technische Universit\"at M\"unchen, Germany (e-mail: manuel.stein@tum.de).}
}

\begin{document}
\maketitle
\begin{abstract}
The problem of determining the achievable sensitivity with digitization exhibiting minimal complexity is addressed. In this case, measurements are exclusively available in hard-limited form. Assessing the achievable sensitivity via the Cram\'{e}r-Rao lower bound requires characterization of the likelihood function, which is intractable for multivariate binary distributions. In this context, the Fisher matrix of the exponential family and a lower bound for arbitrary probabilistic models are discussed. The conservative approximation for Fisher's information matrix rests on a surrogate exponential family distribution connected to the actual data-generating system by two compact equivalences. Without characterizing the likelihood and its support, this probabilistic notion enables designing estimators that consistently achieve the sensitivity as defined by the inverse of the conservative information matrix. For parameter estimation with multivariate binary samples, a quadratic exponential surrogate distribution tames statistical complexity such that a quantitative assessment of an achievable sensitivity level becomes tractable. This fact is exploited for the performance analysis concerning parameter estimation with an array of low-complexity binary sensors in comparison to an ideal system featuring infinite amplitude resolution. Additionally, data-driven assessment by estimating a conservative approximation for the Fisher matrix under recursive binary sampling as implemented in $\Sigma\Delta$-modulating analog-to-digital converters is demonstrated.
\end{abstract}
%
%
%
%
\section{Introduction}
\IEEEPARstart{F}{isher} information is a traditional measure in statistics \cite{Fisher22,Fisher25}. In detection theory, it aids the design of strong hypothesis tests \cite{Wald43,Engle83}, while in estimation theory, it characterizes the performance of efficient inference algorithms operating without bias \cite{Rao45,Cram46}. In particular, in the large sample regime, the inverse of the Fisher matrix is associated with the covariance of maximum-likelihood estimates, that are normally distributed around the true parameters of the data-generating model, see, e.g., \cite{VanDerVaart}. Through this so-called Cram\'er-Rao lower bound (CRLB) and its Bayesian extensions \cite{Trees07}, the Fisher information measure has become an important mathematical tool in statistical signal processing, wireless communication, and information theory. Examples of applications in electrical engineering range from the performance analysis and system design optimization for positioning based on the propagation of radio waves \cite{Patwari05,Antreich11,Yin14}  or visible light \cite{Zhang14}, to radar \cite{Gini02,Bekkerman06} and sonar \cite{Kirubarajan96}, over communication channel estimation \cite{Rice01}, array signal processing \cite{Stoica89,Viberg91,Swindlehurst92}, biomedical imaging \cite{Barbe11} to data-compression \cite{Amari11,Stein14_IPT,Pakrooh15}.
\subsection{Motivation}
While the Fisherian notion of parameter-specific information is useful for theoretical considerations as well as for practical applications, the quantitative evaluation of the information measure can be challenging. In particular, this is the case when the parametric likelihood of the data-generating distribution model is difficult to compute, or taking the expectation of the score function's outer product is intractable. 

As an example, consider the multivariate Gaussian distribution
\begin{align}\label{multi:gauss:model}
p_{\ve{y}}(\ve{y};\ve{\theta})=\frac{\exp{- \frac{1}{2} \ve{y}^{\T} \ve{R}^{-1}_{\ve{y}}(\ve{\theta})\ve{y}}}{\sqrt{ (2\pi)^{M} \det \big(\ve{R}_{\ve{y}}(\ve{\theta})\big) }} 
\end{align}
modeling the stochastic relation between the zero-mean data $\ve{y}\in\ve{\mathcal{Y}},\ve{\mathcal{Y}}\subseteq\fieldR^{M}, M \in\fieldN,$ and the parameters $\ve{\theta}\in\ve{\Theta}, \ve{\Theta}\subset \fieldR^{D}, D \in\fieldN,$ which modulate the covariance matrix $\ve{R}_{\ve{y}}(\ve{\theta})\in\fieldR^{M\times M}$. For this probability law, the likelihood is available in closed-form \eqref{multi:gauss:model}, while the Fisher information measure $\ve{F_y}(\ve{\theta})\in\fieldR^{D\times D}$ exhibits a tractable structure with the $ij$th matrix entry being \cite[pp. 47]{Kay93}
\begin{align}
\left[ \ve{F_y}(\ve{\theta})\right]_{ij} = \frac{1}{2} \tr{ \frac{\partial \ve{R}_{\ve{y}}(\ve{\theta})}{\partial \theta_i} \ve{R}^{-1}_{\ve{y}}(\ve{\theta}) \frac{\partial \ve{R}_{\ve{y}}(\ve{\theta})}{\partial \theta_j} \ve{R}^{-1}_{\ve{y}}(\ve{\theta}) }.
\end{align}
In case the multivariate distribution \eqref{multi:gauss:model} is used as the system model of a sensing apparatus, one implicit assumption is that the acquired data features sufficiently high amplitude resolution, such that quantization effects can be neglected. However, if the digitization of multiple analog sensor outputs has to be performed at fast sampling rates or when large measurement datasets have to be transmitted and stored, high amplitude resolution can be challenging to realize due to instrumentation cost, power consumption, transmission bandwidth, or memory size constraints. In extreme cases, it can be helpful to minimize the amplitude resolution of the signal digitization process or the transmitted and stored data streams. Then the measurement data, featuring one amplitude bit per sample variable, can be modeled as hard-limited observations of the multivariate distribution \eqref{multi:gauss:model}, i.e.,
\begin{align}\label{mapping:sign}
\ve{z}=\sign{\ve{y}},
\end{align}
$\ve{z}\in\ve{\mathcal{Z}}$, where $\operatorname{sign}{(\cdot)}$ is the element-wise signum function such that $\ve{\mathcal{Z}}\subseteq\{-1,1\}^{M}$. After this nonlinear transformation of the multivariate Gaussian data, the likelihood function is
\begin{align}\label{likeli:difficult}
p_{\ve{z}}(\ve{z};\ve{\theta})=\int_{\ve{\mathcal{Y}}(\ve{z})} p_{\ve{y}}(\ve{y};\ve{\theta}) {\rm d}\ve{y},
\end{align}
where $\ve{\mathcal{Y}}(\ve{z})$ denotes the subset in $\fieldR^{M}$, which is mapped to the binary vector $\ve{z}$ by the application of \eqref{mapping:sign}. The integral \eqref{likeli:difficult} is associated with the orthant probabilities, for which analytical expressions with $M>4$ are, in general, an open problem. If one strives to assess the achievable parameter estimation performance with the binary measurements \eqref{mapping:sign} via the CRLB, the required Fisher information matrix $\ve{F_z}(\ve{\theta}) \in\fieldR^{D \times D}$ is obtained by a weighted summation of the score's outer product over the entire discrete support $\ve{\mathcal{Z}}$, i.e.,
\begin{align}\label{fisher:difficult}
\ve{F_z}(\ve{\theta})&=\sum_{\ve{\mathcal{Z}}} \bigg( \frac{\partial  \ln p_{\ve{z}}(\ve{z};\ve{\theta})}{\partial \ve{\theta}}  \bigg)^{\T}  \frac{\partial  \ln p_{\ve{z}}(\ve{z};\ve{\theta})}{\partial \ve{\theta}} p_{\ve{z}}(\ve{z};\ve{\theta}).
\end{align}
So, even if an analytical expression of the likelihood \eqref{likeli:difficult} and its log-derivative\footnote{The notation of derivatives used throughout the manuscript is defined in \eqref{notation:derivatives}.} would be available, calculation of the Fisher information matrix \eqref{fisher:difficult} might still be intractable for moderately large $M$ due to the exponentially growing cardinality $|\ve{\mathcal{Z}}|=2^M$ of the support $\ve{\mathcal{Z}}$. Thus, it is difficult to make a founded and precise estimation-theoretical statement about the level of parameter-specific information contained in sensor measurements acquired by low-complexity binary digitization schemes like \eqref{mapping:sign}.

Concerning the relationship between the Fisher information matrices $\ve{F_y}(\ve{\theta})$ and $\ve{F_z}(\ve{\theta})$, the estimation-theoretical version of the data processing theorem \cite[Lemma 3]{Zamir98} \cite[Proposition 4]{Rioul11} ensures that the parameter-specific information in the measurement data can not increase through hard-limiting, i.e.,
\begin{align}\label{fisher:inequality:data:processing}
\ve{F_y}(\ve{\theta}) \succeq \ve{F_z}(\ve{\theta}).
\end{align}
While this matrix inequality provides qualitative insight, it does, however, not characterize $\ve{F_z}(\ve{\theta})$ quantitatively. As such, the actual amount of parameter-specific information lost during binarization \eqref{mapping:sign} remains unclear. Further, inequality \eqref{fisher:inequality:data:processing} does not provide a guideline on how to make use of the information $\ve{F_z}(\ve{\theta})$ contained in the binary data at the output of the nonlinearity \eqref{mapping:sign}. 

Generalizing to measurements $\ve{u}\in\ve{\mathcal{U}}$ obtained by an arbitrary physical data acquisition process, one notices that the distribution model $p_{\ve{u}}(\ve{u};\ve{\theta})$ and its Fisher information matrix $\ve{F_u}(\ve{\theta})$ can rarely be determined precisely due to the complicated stochastic relationship between the generated data $\ve{u}$ and the underlying parameters $\ve{\theta}$. Then the Fisher information matrix $\ve{F_u}(\ve{\theta})$ needs to be approximated in a data-driven way or by estimating the distribution function, see, e.g., \cite{Spall05,Berisha15}.

The motivation of this article, therefore, is to discuss a particular approach providing matrix inequalities
\begin{align}\label{inequality:pessimistic:fisher}
\ve{F_u}(\ve{\theta}) \succeq \ve{\tilde{F}_u}(\ve{\theta}),
\end{align}
where the conservative information matrix $\ve{\tilde{F}_u}(\ve{\theta})$ on the right-hand side is tractable. Such inequalities are here obtained through the concept of a surrogate distribution model $\tilde{p}_{\ve{u}}(\ve{u};\ve{\theta})$, which can be designed such that $\ve{\tilde{F}_u}(\ve{\theta})$ exhibits a form that allows explicit computation. While the likelihood of the surrogate probabilistic model $\tilde{p}_{\ve{u}}(\ve{u};\ve{\theta})$ remains inaccessible, its score $\frac{\partial \ln \tilde{p}_{\ve{u}}(\ve{u};\ve{\theta})}{\partial \ve{\theta}}$ can be evaluated. Through root-finding algorithms, this enables extracting the parameter-specific information, which by the conservative information matrix $\ve{\tilde{F}_u}(\ve{\theta})$ is guaranteed to be contained in samples from the distribution $p_{\ve{u}}(\ve{u};\ve{\theta})$. Further, the conservative information matrix $\ve{\tilde{F}_u}(\ve{\theta})$ can be approximated by a simple data-driven procedure whenever it is possible to generate large datasets in calibrated environments or by simulations. This allows measuring the amount of parameter-specific information contained in samples from unknown distribution models and optimizing the technical layout of the associated data acquisition process.

In the context of binary sampling theory, the Fisher matrices $\ve{F_y}(\ve{\theta})$ and $\ve{\tilde{F}_z}(\ve{\theta})$ enable quantitatively delimiting the parameter-specific information loss caused by hard-limiting \eqref{mapping:sign} multivariate Gaussian data \eqref{multi:gauss:model}. In electrical engineering, such a conservative sensitivity analysis is of practical relevance as, without designing and running a particular algorithm, it allows evaluating the performance achievable through statistical processing of data acquired by sensing systems with minimal analog-to-digital (A/D) conversion complexity or of digital measurements stored on compact hard-limiting memory devices. 
\subsection{Related Work}
Classical discussions dealing with the analysis of nonlinear systems are \cite{Volterra87,Wiener58,Brilliant58,Volterra59} and concentrate on a deterministic characterization of the output in terms of a series of polynomial integral operations applied to the input. Under a probabilistic perspective, \cite{Kac47,Meyer54} investigate the distribution function of the nonlinear output, while \cite{Rice44,Bennett48,Middleton48,Bussgang52,Price58,Vleck66,Hinich67,Baum69,Hall69} focus on average statistical properties, like spectra or cross- and autocorrelation functions. By considering statistical measures of signal quality such as signal-to-noise ratio (SNR) and distortion-to-signal power ratio, the manuscripts \cite{Davenport53,Rowe82} can be considered as early work on information flow in nonlinear systems. In modern communication theory and the context of Shannon's information measure, this discussion finds continuation by contributions such as \cite{Zhang12,Mezghani12}. In contemporary statistical signal processing and the context of the Fisher information measure, \cite{HostMadsen00,BarShalom02,Ribeiro06,Balkan10,Mezghani10} are examples that attest to the relevance of characterizing the information loss induced by nonlinear systems, in particular, coarse quantization functions like \eqref{mapping:sign}.

Fundamental properties of Fisher information are discussed in \cite{StamPhD59}, while \cite{Boek77} considers a more general definition, and \cite{Cohen68} establishes convexity of the measure. For the role of Fisher information in the interplay between estimation and information theory, see, e.g. \cite{Stam59,Blachman65,Costa84,Dembo91,Zamir98,Rioul11,Lutwak12,Guo13}. Deriving the distribution, which provides minimum Fisher information under moment-constraints, is considered in \cite{Zivo97}, while \cite{UhrKli95,Berch09} treat this aspect under a restricted support. For models with independent additive noise, \cite{Stoica11,Park13} show that assuming Gaussian noise minimizes Fisher information and, therefore, maximizes the CRLB. For performance analysis of signal parameter estimation in nonlinear sensing applications, \cite{Stein14,Stein15} generalize the discussion to observation models with dependent non-additive noise by lower bounding Fisher information and identifying connections to the independent additive Gaussian models commonly used in the electrical engineering sciences.

A discourse on lower bounds for the Fisher information measure is also available through literature in statistics. There the problem is, to the best of the author's knowledge, formulated the first time in \cite{Sankaran64} as a side-product of a discussion on lower bounds for the variance of unbiased estimates. Subsequently, a seminal treatment of lower bounds for the Fisher information measure based on moments and orthogonal statistics is provided by \cite{Jarrett84}, while \cite{Zografos94} states a generalized lower bound in matrix form and shows that exponential family distributions minimize it. In the absence of an early discussion stressing the practical relevance of such conservative approximations for Fisher's information measure, these works in statistics have not received scientific attention proportional to their technical importance.
\subsection{Contribution}
This article intends to provide such a discussion by considering the theoretical topic of lower bounds for the Fisher information matrix while emphasizing practical applicability in the context of electrical engineering. For this purpose, a general lower bound for Fisher's information measure (stated in \cite[Sec. 2]{Zografos94} without proof) is derived and the achievability of the sensitivity level associated with its inverse is established. The information bound is used to quantitatively assess the estimation performance, which is guaranteed through likelihood-oriented statistical processing of measurements acquired by binary sampling schemes. In particular, the inference of the signal-to-noise (SNR) and direction-of-arrival (DOA) parameters of a narrow-band wireless source with a large-scale array of binary sensors is considered aside with an ideal reference system featuring infinite digital amplitude resolution. Additionally, for a noisy single-sensor signal with unknown mean (DC offset) and variance (AC power), the parameter-specific information loss resulting from recursive binary sampling schemes with different error feedback weights and oversampling factors is determined by measuring a conservative Fisher information matrix via Monte-Carlo simulations. The presented analysis corroborates the potential of low-complexity binary sensing technology for applications where physical model parameters characterizing a noisy analog received signal have to be digitally inferred with high precision.

\subsection{Outline}
The discourse starts by analyzing the Fisher information matrix of multivariate exponential family distributions, resulting in an identity connecting the information matrix to a weighted sum of derivatives of the expected sufficient statistics. The identity suggests that, under certain conditions, the exact computation of the Fisher information matrix for a broad class of practically relevant distributions is rather simple. This renders the probabilistic framework of the exponential family particularly suitable for tractable approximations of the Fisher information matrix under arbitrary data models. Consequently, the discussion is generalized to a generic probability distribution $p_{\ve{u}}(\ve{u};\ve{\theta})$, which is not necessarily part of the exponential family. The actual probabilistic model $p_{\ve{u}}(\ve{u};\ve{\theta})$ stands in relation to a surrogate distribution $\tilde{p}_{\ve{u}}(\ve{u};\ve{\theta})$ framed by the exponential family with sufficient statistics $\tilde{\ve{\phi}}(\ve{u})$. The link between the two probability laws is that the parametric mean and covariance of the surrogate statistics $\tilde{\ve{\phi}}(\ve{u})$ are equivalent under both distributions\footnote{In the following, two such distributions are considered as equivalent to each other with respect to $\tilde{\ve{\phi}}(\ve{u})$. This denomination is intended to emphasize that here no less preference is given to the exponential surrogate model than to the original system.}. The covariance inequality \cite[Sec. 1.2.1]{Trees07} then ensures that the information matrix $\ve{\tilde{F}_u}(\ve{\theta})$ of the equivalent exponential family $\tilde{p}_{\ve{u}}(\ve{u};\ve{\theta})$ is always dominated by the Fisher matrix $\ve{F_u}(\ve{\theta})$ of the original system $p_{\ve{u}}(\ve{u};\ve{\theta})$ and, as such, constitutes a lower bound in matrix form. The theoretical concept of the equivalent exponential family $\tilde{p}_{\ve{u}}(\ve{u};\ve{\theta})$ is useful in practice as the statistics $\tilde{\ve{\phi}}(\ve{u})$ can be chosen such that a quantitative evaluation of the information matrix $\ve{\tilde{F}_u}(\ve{\theta})$ becomes tractable without explicit characterization of the likelihood functions $p_{\ve{u}}(\ve{u};\ve{\theta})$ or $\tilde{p}_{\ve{u}}(\ve{u};\ve{\theta})$. Therefore, the Fisher matrix $\ve{F_u}(\ve{\theta})$ of the data-generating model can be conservatively explored through tractable versions of $\ve{\tilde{F}_u}(\ve{\theta})$ based exclusively on the mean derivatives and covariance of the user-defined statistics $\tilde{\ve{\phi}}(\ve{u})$.

For the extraction of the parameter-specific information $\ve{\tilde{F}_u}(\ve{\theta})$, which is guaranteed to be contained in samples from ${p}_{\ve{u}}(\ve{u};\ve{\theta})$, it is established that the error covariance characterized by the inverse of $\ve{\tilde{F}_u}(\ve{\theta})$ can be achieved through estimators calculated based on the score function of $\tilde{p}_{\ve{u}}(\ve{u};\ve{\theta})$. Reformulation of such inference procedures shows that the equivalent exponential family distribution $\tilde{p}_{\ve{u}}(\ve{u};\ve{\theta})$ connects Pearson's method of moments \cite{Pearson94,Hansen82} to Fisher's technique of maximum-likelihood estimation \cite{Fisher22}. In an asymptotic sense, the conservative information matrix $\ve{\tilde{F}_u}(\ve{\theta})$ enables interpreting the equivalent exponential family distribution $\tilde{p}_{\ve{u}}(\ve{u};\ve{\theta})$ as a particular multivariate Gaussian modeling framework.

To demonstrate application in electrical engineering, the conservative information matrix is used to determine the achievable estimation sensitivity with sensor data obtained through binary sampling. On the one hand, signal parameter estimation with hard-limited measurements produced by an array of receivers featuring low-complexity $1$-bit analog-to-digital (A/D) conversion is considered. In particular, the effect of increasing the number of array elements onto the parameter-specific information flow from the analog sensor outputs to the digital processing unit is investigated. On the other hand, the use of the information bound in scenarios where the conservative information matrix $\ve{\tilde{F}_z}(\ve{\theta})$ is challenging to derive mathematically is outlined. To this end, the task of parameter estimation from digital measurements obtained via recursive binary sampling schemes with oversampling is examined. The Fisher information matrix of an equivalent quadratic exponential family is measured by Monte-Carlo simulations of the nonlinear output $\ve{z}$ under an input model $p_{\ve{y}}(\ve{y};\ve{\theta})$ with fixed parameters $\ve{\theta}$. This provides a quantitative data-driven assessment of the parameter-specific information contained in the samples drawn from the unknown multivariate distribution $p_{\ve{z}}(\ve{z};\ve{\theta})$ characterizing the binary output of the nonlinear recursive data acquisition. In particular, it is illustrated how this approach enables identifying error feedback designs for $\Sigma\Delta$-modulation, which maximize the parameter-specific information in the acquired data stream. 

Note that a significant portion of the discussion is part of the author's doctoral thesis \cite{SteinPhD}. A discourse regarding the application of the conservative Fisher information matrix $\ve{\tilde{F}_z}(\ve{\theta})$ in the context of $1$-bit DOA parameter estimation is given in the conference contribution \cite{SteinWSA16}. Lower bounding the Fisher information measure of probabilistic models with a scalar parameter and complementary engineering applications in the context of instrumentation and measurement are outlined in \cite{SteinI2MTC16,SteinTIM2018}.
\section{Fisher Information and the Exponential Family}\label{sec:fisher:info:exp:family}
The family of probability measures, denoted by a density or mass function $p_{\ve{u}}(\ve{u};\ve{\theta})$, modeling continuous or discrete $M$-variate random data $\ve{u}\in\ve{\mathcal{U}}$ with deterministic unknown parameters $\ve{\theta}\in\ve{\Theta}$ is considered. $\ve{\mathcal{U}}$ denotes the support of the probability law $p_{\ve{u}}(\ve{u};\ve{\theta})$ while $\ve{\Theta}\subset\fieldR^D$ represents the parameter space of $\ve{\theta}$. Throughout the discussion, it is assumed that all distribution models exhibit regularity and are twice-differentiable concerning their parameters, implying that the support $\ve{\mathcal{U}}$ is independent of the model parameters $\ve{\theta}$. Following the notational convention 
\begin{align}\label{notation:derivatives}
\left[ \frac{\partial \ve{f}(\ve{x})}{\partial \ve{x}} \right]_{ij} = \frac{\partial {f}_i(\ve{x})}{\partial x_j},
\end{align}
the Fisher information matrix of such distributions is
\begin{align}\label{def:fisher:matrix}
\ve{F_u}(\ve{\theta})&=\exdi{\ve{u};\ve{\theta}}{ \bigg( \frac{\partial \ln p_{\ve{u}}(\ve{u};\ve{\theta})}{\partial \ve{\theta}} \bigg)^{\T} \frac{\partial \ln p_{\ve{u}}(\ve{u};\ve{\theta})}{\partial \ve{\theta}} }\\
&= - \exdi{\ve{u};\ve{\theta}}{\frac{\partial^2 \ln {p}_{\ve{u}}(\ve{u};\ve{\theta})}{\partial \ve{\theta}^2} },
\end{align}
$\ve{F_u}(\ve{\theta})\in\fieldR^{D \times D}$. Note that the discussion is restricted to distributions $p_{\ve{u}}(\ve{u};\ve{\theta})$ with locally identifiable parameters such that the Fisher information matrix is non-singular \cite{Rothenberg71}.
\subsection{Distribution Model - Exponential Family}
The exponential family are the probability distributions for which the likelihood can be written as
\begin{align}\label{exp:family:vec}
p_{\ve{u}}(\ve{u};\ve{\theta})=\exp{\sum_{l=1}^{L} w_l(\ve{\theta}) \phi_l(\ve{u}) - \lambda(\ve{\theta})+{\kappa(\ve{u})}},
\end{align}
where $w_l(\ve{\theta}): \fieldR^{D}\to\fieldR$ is the $l$th statistical weight\footnote{In the statistics literature usually the term ``natural parameter" is used for $w_l(\ve{\theta})$. Trying to make a clear distinction between the elementary mathematical components of the probabilistic data model $p_{\ve{u}}(\ve{u};\ve{\theta})$ and its parameters $\ve{\theta}$, which in electrical engineering are usually associated with a physical/natural phenomena, the term ``statistical weight" for $w_l(\ve{\theta})$ is used.}, $\phi_l(\ve{u}): \ve{\mathcal{U}}\to\fieldR$ the $l$th sufficient statistic, $\lambda(\ve{\theta}): \fieldR^{D}\to\fieldR$ the log-normalizer and $\kappa(\ve{u}): \ve{\mathcal{U}}\to\fieldR$ the carrier measure. 

The log-likelihood of an exponential family distribution is
\begin{align}
\ln p_{\ve{u}}(\ve{u};\ve{\theta}) &= \sum_{l=1}^{L} w_l(\ve{\theta}) \phi_l(\ve{u}) - \lambda(\ve{\theta})+{\kappa(\ve{u})},
\end{align}
while its score function has the structure
\begin{align}\label{exp:score:vec}
\frac{\partial \ln p_{\ve{u}}(\ve{u};\ve{\theta})}{\partial \ve{\theta}} &= \sum_{l=1}^{L} \frac{\partial w_l(\ve{\theta})}{\partial \ve{\theta}} \phi_l(\ve{u}) - \frac{\partial\lambda(\ve{\theta})}{\partial \ve{\theta}}.
\end{align}
Due to regularity
\begin{align}\label{exp:score:vec:zero}
\exdi{\ve{u};\ve{\theta}}{\frac{\partial \ln p_{\ve{u}}(\ve{u};\ve{\theta})}{\partial \ve{\theta}}}=\ve{0}^{\T},
\end{align}
where $\exdi{\ve{u};\ve{\theta}}{\cdot}$ denotes the expectation regarding $p_{\ve{u}}(\ve{u};\ve{\theta})$ and, depending on the context, $\ve{0}$ is a vector or a matrix with all entries equal to zero. Using \eqref{exp:score:vec} in \eqref{exp:score:vec:zero} provides
\begin{align}\label{identity:log:normalizer}
\sum_{l=1}^{L} \frac{\partial w_l(\ve{\theta})}{\partial \ve{\theta}} \exdi{\ve{u};\ve{\theta}}{\phi_l(\ve{u})} = \frac{\partial\lambda(\ve{\theta})}{\partial \ve{\theta}}.
\end{align}

Note that, through the sufficient statistics, the distributions featuring an exponential family structure \eqref{exp:family:vec} have the useful property that the information contained in an arbitrary number of samples can be represented without loss in $L$ values \cite{Koopman36,Pitman36}. Further, given a set of moment constraints, the probability distribution maximizing Shannon's measure of entropy resides in the exponential family \cite{Jaynes57,Kapur92,Cover06}.
\subsection{Fisher Information Matrix of the Exponential Family}
Substituting one of the score functions in \eqref{def:fisher:matrix} by \eqref{exp:score:vec}, gives
\begin{align}
\ve{F_u}(\ve{\theta})&=\exdi{\ve{u};\ve{\theta}}{ \bigg( \frac{\partial \ln p_{\ve{u}}(\ve{u};\ve{\theta})}{\partial \ve{\theta}} \bigg)^{\T} \sum_{l=1}^{L} \frac{\partial w_l(\ve{\theta})}{\partial \ve{\theta}} \phi_l(\ve{u})  }\notag\\
&-\exdi{\ve{u};\ve{\theta}}{ \bigg( \frac{\partial \ln p_{\ve{u}}(\ve{u};\ve{\theta})}{\partial \ve{\theta}} \bigg)^{\T} \frac{\partial\lambda(\ve{\theta})}{\partial \ve{\theta}} },
\end{align}
such that with regularity \eqref{exp:score:vec:zero}, it follows that
\begin{align}\label{fish:identity:vec}
\ve{F_u}(\ve{\theta})&=  \sum_{l=1}^{L}  \bigg( \exdi{\ve{u};\ve{\theta}}{ \frac{\partial \ln p_{\ve{u}}(\ve{u};\ve{\theta})}{\partial \ve{\theta}} \phi_l(\ve{u}) } \bigg)^{\T} \frac{\partial w_l(\ve{\theta})}{\partial \ve{\theta}} \notag\\
&=\sum_{l=1}^{L}   \bigg(\int_{\ve{\mathcal{U}}} \frac{\partial  p_{\ve{u}}(\ve{u};\ve{\theta})}{\partial \ve{\theta}} \phi_l(\ve{u})  {\rm d} \ve{u} \bigg)^{\T}   \frac{\partial w_l(\ve{\theta})}{\partial \ve{\theta}}\notag\\
&=\sum_{l=1}^{L}   \bigg( \frac{\partial \exdi{\ve{u};\ve{\theta}}{ \phi_l(\ve{u}) } }{\partial \ve{\theta}}  \bigg)^{\T}   \frac{\partial w_l(\ve{\theta})}{\partial \ve{\theta}}.
\end{align}
Defining a vector with the sufficient statistics
\begin{align}
\ve{\phi}(\ve{u})=\begin{bmatrix} \phi_1(\ve{u}) &\phi_2(\ve{u}) &\ldots &\phi_L(\ve{u})\end{bmatrix}^{\T},
\end{align}
$\ve{\phi}(\ve{u}) \in \fieldR^{L}$, its expected value
\begin{align}
\ve{\mu}_{\ve{\phi}}(\ve{\theta}) = \exdi{\ve{u};\ve{\theta}}{\ve{\phi}(\ve{u})},
\end{align}
$\ve{\mu}_{\ve{\phi}}(\ve{\theta}) \in \fieldR^{L}$, and a vector with the statistical weights 
\begin{align}
\ve{w}(\ve{\theta})=\begin{bmatrix} w_1(\ve{\theta}) &w_2(\ve{\theta}) &\ldots &w_L(\ve{\theta})\end{bmatrix}^{\T},
\end{align}
$\ve{w}(\ve{\theta}) \in \fieldR^{L}$, Fisher's information measure \eqref{fish:identity:vec} can be stated in compact form
\begin{align}\label{fish:identity:vec:compact}
\ve{F_u}(\ve{\theta})&= \bigg( \frac{\partial \ve{\mu}_{\ve{\phi}}(\ve{\theta})}{\partial \ve{\theta}} \bigg)^{\T} \frac{\partial \ve{w}(\ve{\theta})}{\partial \ve{\theta}}.
\end{align}
\subsection{Example - Multivariate Zero-Mean Gaussian Distribution}
To illustrate application of the identity \eqref{fish:identity:vec:compact}, the multivariate Gaussian model \eqref{multi:gauss:model} is considered. Its sufficient statistics are 
\begin{align}\label{sufficient:stat:gaussian:zero:mean}
\ve{\phi}(\ve{y})=\vec{\ve{y}\ve{y}^{\T}}
\end{align}
and the statistical weights
\begin{align}
\ve{w}(\ve{\theta})=-\frac{1}{2}\vec{\ve{R}_{\ve{y}}^{-1}(\ve{\theta})},
\end{align}
where the covariance matrix is defined
\begin{align}\label{definition:gauss:covariance:matrix}
\ve{R}_{\ve{y}}(\ve{\theta})=\exdi{\ve{y};\ve{\theta}}{\ve{y}\ve{y}^{\T}},
\end{align}
$\ve{R}_{\ve{y}}(\ve{\theta})  \in \fieldR^{M \times M}$ and the matrix operator $\vec{\ve{B}}$ provides the columns of the matrix $\ve{B}$ stacked in one column vector. The parametric expectancy of the sufficient statistics \eqref{sufficient:stat:gaussian:zero:mean} is 
\begin{align}
\ve{\mu}_{\ve{\phi}}(\ve{\theta})=\exdi{\ve{y};\ve{\theta}}{\ve{\phi}(\ve{y})}=\vec{\ve{R}_{\ve{y}}(\ve{\theta})},
\end{align}
such that the derivative
\begin{align}
\frac{\partial \ve{\mu}_{\ve{\phi}}(\ve{\theta})}{\partial \ve{\theta}}&=\begin{bmatrix} \frac{\partial \ve{\mu}_{\ve{\phi}}(\ve{\theta})}{\partial \theta_1} & \frac{\partial \ve{\mu}_{\ve{\phi}}(\ve{\theta})}{\partial \theta_2} &\ldots &\frac{\partial \ve{\mu}_{\ve{\phi}}(\ve{\theta})}{\partial \theta_D} \end{bmatrix}
\end{align}
is a matrix with the $d$th column
\begin{align}
\frac{\partial \ve{\mu}_{\ve{\phi}}(\ve{\theta})}{\partial \theta_d} = \vec{ \frac{\partial \ve{R}_{\ve{y}}(\ve{\theta})}{\partial \theta_d} }.
\end{align}
The derivative of the statistical weights is
\begin{align}
\frac{\partial \ve{w}(\ve{\theta}) }{\partial \ve{\theta} }&=\begin{bmatrix} \frac{\partial \ve{w}(\ve{\theta})}{\partial \theta_1} & \frac{\partial \ve{w}(\ve{\theta}) }{\partial \theta_2} &\ldots &\frac{\partial \ve{w}(\ve{\theta})} {\partial \theta_D} \end{bmatrix},
\end{align}
with the $d$th column being
\begin{align}
 \frac{\partial \ve{w}(\ve{\theta})}{\partial \theta_d} = \frac{1}{2}\vec{ \ve{R}_{\ve{y}}^{-1}(\ve{\theta}) \frac{\partial \ve{R}_{\ve{y}}(\ve{\theta})}{\partial \theta_d} \ve{R}_{\ve{y}}^{-1}(\ve{\theta})}.
\end{align}
With \eqref{fish:identity:vec:compact}, it follows that the $i$th row and $j$th column entry of the Fisher information matrix is
\begin{align}\label{fisher:matrix:entries:gauss:simple}
&\left[ \ve{F_y}(\ve{\theta}) \right]_{ij}=\notag\\
&=\frac{1}{2} \vec{ \frac{\partial \ve{R}_{\ve{y}}(\ve{\theta})}{\partial \theta_i} }^{\T} \vec{ \ve{R}_{\ve{y}}^{-1}(\ve{\theta}) \frac{\partial \ve{R}_{\ve{y}}(\ve{\theta})}{\partial \theta_j} \ve{R}_{\ve{y}}^{-1}(\ve{\theta})}\notag\\
&=\frac{1}{2} \tr{ \frac{\partial \ve{R}_{\ve{y}}(\ve{\theta})}{\partial \theta_i} \ve{R}_{\ve{y}}^{-1}(\ve{\theta}) \frac{\partial \ve{R}_{\ve{y}}(\ve{\theta})}{\partial \theta_j} \ve{R}_{\ve{y}}^{-1}(\ve{\theta})},
\end{align}
where the last step follows from the properties of the vec-operator and $\tr{\ve{B}}$ denotes the trace of the square matrix $\ve{B}$. The fact that \eqref{fisher:matrix:entries:gauss:simple} characterizes the entries of the Fisher information matrix \cite[pp. 47]{Kay93} of the multivariate Gaussian distribution model \eqref{multi:gauss:model} shows that the identity \eqref{fish:identity:vec:compact} allows to shorten the usual calculation of \eqref{def:fisher:matrix} along the score, see, e.g., \cite[pp. 73]{Kay93}.
\section{Lower Bound for the Fisher Information Matrix}\label{sec:fisher:info:lower:bound}
If the parametric distribution function $p_{\ve{u}}(\ve{u};\ve{\theta})$ belongs to the exponential family \eqref{exp:family:vec} and its statistical weights $\ve{w}(\ve{\theta})$ as well as the expectancy of the sufficient statistics $\ve{\mu}_{\ve{\phi}}(\ve{\theta})$ are known in analytic form, \eqref{fish:identity:vec:compact} shows that the Fisher information matrix \eqref{def:fisher:matrix} can be computed by taking derivatives and evaluating a matrix product. In such cases, it is not required to explicitly specify all the components of the likelihood \eqref{exp:family:vec} and take the expectation \eqref{def:fisher:matrix} of the score function's outer product. In cases where the probability distribution function $p_{\ve{u}}(\ve{u};\ve{\theta})$ does not belong to the exponential family \eqref{exp:family:vec}, the parametric mean of the sufficient statistics $\ve{\mu}_{\ve{\phi}}(\ve{\theta})$ is unknown, or the statistical weights $\ve{w}(\ve{\theta})$ are not available, \eqref{fish:identity:vec:compact} can not be used to determine the Fisher matrix. Nevertheless, the probabilistic framework of the exponential family and the compact structure of its Fisher information matrix provide a gentle foundation for constructing mathematically tractable or data-driven approximations for the Fisher information matrix of arbitrary distribution models. 

\subsection{Approach - Equivalent Exponential Family Distribution}
To this end, in the following $p_{\ve{u}}(\ve{u};\ve{\theta})$ is considered to be a generic $M$-variate probability law with support $\ve{\mathcal{U}}$. Additionally, an $M$-variate surrogate distribution $\tilde{p}_{\ve{u}}(\ve{u};\ve{\theta})$, with support $\ve{\tilde{\mathcal{U}}}$ and exponential family factorization \eqref{exp:family:vec}, which features the $\tilde{L} \geq D$ user-defined sufficient statistics
\begin{align}\label{definition:auxiliary:statistics}
\ve{\tilde{\phi}}(\ve{u}): \ve{\mathcal{U}}\cup\ve{\tilde{\mathcal{U}}} \to \fieldR^{\tilde{L}}
\end{align}
is defined. The two probabilistic models $\tilde{p}_{\ve{u}}(\ve{u};\ve{\theta})$ and $p_{\ve{u}}(\ve{u};\ve{\theta})$ are solely connected by the equivalences 
\begin{align}\label{equivalence:mean}
\ve{\mu}_{\ve{\tilde{\phi}}}(\ve{\theta})=\exdi{\ve{\tilde{u}};\ve{\theta}}{  \ve{\tilde{\phi}}(\ve{u}) }=\exdi{\ve{u};\ve{\theta}}{  \ve{\tilde{\phi}}(\ve{u}) }
\end{align}
and 
\begin{align}\label{equivalence:covariance}
\ve{R}_{\ve{\tilde{\phi}}}(\ve{\theta})&=\exdi{\ve{\tilde{u}};\ve{\theta}}{\big(\ve{\tilde{\phi}}(\ve{u})-\ve{\mu}_{\ve{\tilde{\phi}}}(\ve{\theta})\big)\big(\ve{\tilde{\phi}}(\ve{u})-\ve{\mu}_{\ve{\tilde{\phi}}}(\ve{\theta})\big)^{\T}  }\notag\\
&=\exdi{\ve{u};\ve{\theta}}{\big(\ve{\tilde{\phi}}(\ve{u})-\ve{\mu}_{\ve{\tilde{\phi}}}(\ve{\theta})\big)\big(\ve{\tilde{\phi}}(\ve{u})-\ve{\mu}_{\ve{\tilde{\phi}}}(\ve{\theta})\big)^{\T}},
\end{align}
where $\exdi{\ve{\tilde{u}};\ve{\theta}}{\cdot}$ denotes the expectation with respect to the surrogate distribution $\tilde{p}_{\ve{u}}(\ve{u};\ve{\theta})$. The equivalence of the statistics' mean \eqref{equivalence:mean} under both distributions is the moment constraint found in the context of maximum-entropy distributions, while \eqref{equivalence:covariance} is an additional constraint on the covariance of the statistics \eqref{definition:auxiliary:statistics}. Note that \eqref{equivalence:mean} and \eqref{equivalence:covariance} allow that the support $\ve{\mathcal{\tilde{U}}}$ of the surrogate data model $\tilde{p}_{\ve{u}}(\ve{u};\ve{\theta})$ is different from the support $\ve{\mathcal{U}}$ of the original distribution function $p_{\ve{u}}(\ve{u};\ve{\theta})$. In the following, the surrogate distribution $\tilde{p}_{\ve{u}}(\ve{u};\ve{\theta})$ is referred to as an equivalent exponential family with respect to the original data model $p_{\ve{u}}(\ve{u};\ve{\theta})$. Further it is assumed that the statistics \eqref{definition:auxiliary:statistics} are designed by the user such that the covariance matrix \eqref{equivalence:covariance} has full rank\footnote{For statistics $\ve{\tilde{\phi}}'(\ve{u}) \in \fieldR^{\tilde{L}'}$, with rank-deficient covariance $\ve{R}_{\ve{\tilde{\phi}}'}(\ve{\theta}) \in \fieldR^{\tilde{L}' \times \tilde{L}'}$, one can modify $\ve{\tilde{\phi}}'(\ve{u})$ by projecting onto the $\tilde{L}$-dimensional span ($\tilde{L}'> \tilde{L}$) of the covariance $\ve{R}_{\ve{\tilde{\phi}}'}(\ve{\theta})$ or by eliminating some of the statistics.}.

With the surrogate model $\tilde{p}_{\ve{u}}(\ve{u};\ve{\theta})$ being defined in the exponential family \eqref{exp:family:vec}, its score is
\begin{align}\label{definition:exponential:replacement:score}
\frac{\partial \ln \tilde{p}_{\ve{u}}(\ve{u};\ve{\theta})}{\partial \ve{\theta}}= \ve{\tilde{\phi}}^{\T}(\ve{u})\frac{\partial \ve{\tilde{w}}(\ve{\theta})}{\partial \ve{\theta}} - \frac{\partial\tilde{\lambda}(\ve{\theta})}{\partial \ve{\theta}}.
\end{align}
Like in \eqref{identity:log:normalizer}, due to regularity \eqref{exp:score:vec:zero},
\begin{align}
\frac{\partial\tilde{\lambda}(\ve{\theta})}{\partial \ve{\theta}} = \ve{\mu}_{\ve{\tilde{\phi}}}^{\T}(\ve{\theta})\frac{\partial \ve{\tilde{w}}(\ve{\theta})}{\partial \ve{\theta}},
\end{align}
such that the auxiliary score function can be written as
\begin{align}\label{exponential:replacement:score:wtilde}
\frac{\partial \ln \tilde{p}_{\ve{u}}(\ve{u};\ve{\theta})}{\partial \ve{\theta}}= \big(\ve{\tilde{\phi}}(\ve{u}) - \ve{\mu}_{\ve{\tilde{\phi}}}(\ve{\theta})  \big)^{\T} \frac{\partial \ve{\tilde{w}}(\ve{\theta})}{\partial \ve{\theta}}.
\end{align}
\subsection{Result - Fisher Information Matrix Lower Bound}
Using the covariance inequality \cite[Sec. 1.2.1]{Trees07}, it is possible to show that the equivalent exponential distribution $\tilde{p}_{\ve{u}}(\ve{u};\ve{\theta})$ is a conservative surrogate data model concerning Fisher information.

To this end, note that, due to regularity \eqref{exp:score:vec:zero}, it holds that
\begin{align}\label{fish:expo:info:ident1}
&\exdi{\ve{u};\ve{\theta}}{\bigg( \frac{\partial \ln p_{\ve{u}}(\ve{u};\ve{\theta})}{\partial \ve{\theta}} \bigg)^{\T}  \frac{\partial \ln \tilde{p}_{\ve{u}}(\ve{u};\ve{\theta})}{\partial \ve{\theta}} }=\notag\\
&=\exdi{\ve{u};\ve{\theta}}{ \bigg( \frac{\partial \ln p_{\ve{u}}(\ve{u};\ve{\theta})}{\partial \ve{\theta}} \bigg)^{\T} \big(\ve{\tilde{\phi}}(\ve{u}) - \ve{\mu}_{\ve{\tilde{\phi}}}(\ve{\theta})  \big)^{\T} \frac{\partial \ve{\tilde{w}}(\ve{\theta})}{\partial \ve{\theta}}}\notag\\
&=\bigg( \int_{\ve{\mathcal{U}}} \ve{\tilde{\phi}}(\ve{u}) \frac{\partial p_{\ve{u}}(\ve{u};\ve{\theta})}{\partial \ve{\theta}}  {\rm d}\ve{u} \bigg)^{\T} \frac{\partial \ve{\tilde{w}}(\ve{\theta})}{\partial \ve{\theta}}\notag\\
&= \bigg(\frac{\partial \ve{\mu}_{\ve{\tilde{\phi}}} (\ve{\theta}) }{\partial \ve{\theta}} \bigg)^{\T} \frac{\partial \ve{\tilde{w}}(\ve{\theta})}{\partial \ve{\theta}}\notag\\
&=\ve{\tilde{F}_u}(\ve{\theta}),
\end{align}
where the last step follows from the identity \eqref{fish:identity:vec:compact}. With the score function \eqref{exponential:replacement:score:wtilde} and the covariance constraint \eqref{equivalence:covariance}, one obtains
\begin{align}\label{fish:expo:info:ident2}
\ve{\tilde{F}}_{\ve{u}}(\ve{\theta})&=\exdi{\ve{\tilde{u}};\ve{\theta}}{\bigg( \frac{\partial \ln \tilde{p}_{\ve{u}}(\ve{u};\ve{\theta})}{\partial \ve{\theta}} \bigg)^{\T}  \frac{\partial \ln \tilde{p}_{\ve{u}}(\ve{u};\ve{\theta})}{\partial \ve{\theta}} } \notag\\
&=\exdi{\ve{u};\ve{\theta}}{\bigg( \frac{\partial \ln \tilde{p}_{\ve{u}}(\ve{u};\ve{\theta})}{\partial \ve{\theta}} \bigg)^{\T}  \frac{\partial \ln \tilde{p}_{\ve{u}}(\ve{u};\ve{\theta})}{\partial \ve{\theta}}}.
\end{align}
By the covariance inequality \cite[Sec. 1.2.1]{Trees07}, it holds that
\begin{align}\label{eq:fisher:expo:replacement}
\ve{F}_{\ve{u}}(\ve{\theta})&=\exdi{\ve{u};\ve{\theta}}{ \bigg( \frac{\partial \ln p_{\ve{u}}(\ve{u};\ve{\theta})}{\partial \ve{\theta}} \bigg)^{\T} \frac{\partial \ln p_{\ve{u}}(\ve{u};\ve{\theta})}{\partial \ve{\theta}} }\notag\\
&\succeq \exdi{\ve{u};\ve{\theta}}{\bigg( \frac{\partial \ln p_{\ve{u}}(\ve{u};\ve{\theta})}{\partial \ve{\theta}} \bigg)^{\T}  \frac{\partial \ln \tilde{p}_{\ve{u}}(\ve{u};\ve{\theta})}{\partial \ve{\theta}} } \notag\\
&\,\,\,\,\cdot\exdi{\ve{u};\ve{\theta}}{ \bigg( \frac{\partial \ln \tilde{p}_{\ve{u}}(\ve{u};\ve{\theta})}{\partial \ve{\theta}} \bigg)^{\T} \frac{\partial \ln \tilde{p}_{\ve{u}}(\ve{u};\ve{\theta})}{\partial \ve{\theta}} }^{-1}\notag\\
&\,\,\,\,\cdot\exdi{\ve{u};\ve{\theta}}{  \bigg(\frac{\partial \ln \tilde{p}_{\ve{u}}(\ve{u};\ve{\theta})}{\partial \ve{\theta}} \bigg)^{\T}  \frac{\partial \ln p_{\ve{u}}(\ve{u};\ve{\theta})}{\partial \ve{\theta}} },
\end{align}
where $\ve{B}\succeq\ve{B'}$ denotes that $\ve{B}-\ve{B'}$ is a positive semidefinite matrix. Therefore, substituting \eqref{fish:expo:info:ident1} and \eqref{fish:expo:info:ident2} in \eqref{eq:fisher:expo:replacement}, while using the symmetry of Fisher's information matrix, yields
\begin{align}\label{fisher:information:inequality:exp:rep}
\ve{F}_{\ve{u}}(\ve{\theta})&\succeq \ve{\tilde{F}_u}(\ve{\theta}).
\end{align}
Inequality \eqref{fisher:information:inequality:exp:rep} shows that the original Fisher information matrix \eqref{def:fisher:matrix} dominates the Fisher information matrix \eqref{fish:expo:info:ident2} of any equivalent exponential family. Interpreting the sufficient statistics \eqref{definition:auxiliary:statistics} of the surrogate model as a processing step, this is in line with the qualitative matrix inequality \eqref{fisher:inequality:data:processing}. The inequality \eqref{fisher:information:inequality:exp:rep}, however, has the advantage that its right-hand side can be evaluated quantitatively through \eqref{equivalence:mean} and \eqref{equivalence:covariance}.

To see this, note that with the score \eqref{exponential:replacement:score:wtilde}, one has
\begin{align}
\ve{\tilde{F}_u}(\ve{\theta})&=\exdi{\ve{\tilde{u}};\ve{\theta}}{\bigg( \frac{\partial \ln \tilde{p}_{\ve{u}}(\ve{u};\ve{\theta})}{\partial \ve{\theta}} \bigg)^{\T}  \frac{\partial \ln \tilde{p}_{\ve{u}}(\ve{u};\ve{\theta})}{\partial \ve{\theta}} }\notag\\
&= \bigg( \frac{\partial \ve{\tilde{w}}(\ve{\theta})}{\partial \ve{\theta}} \bigg)^{\T} \ve{R}_{\ve{\tilde{\phi}}}(\ve{\theta}) \frac{\partial \ve{\tilde{w}}(\ve{\theta})}{\partial \ve{\theta}}.
\end{align}
Comparing to \eqref{fish:identity:vec:compact}, it needs to hold that
\begin{align}\label{append:fish:bound:tightness:equality1}
\ve{\tilde{F}_u}(\ve{\theta})&=\bigg( \frac{\partial \ve{\tilde{w}}(\ve{\theta})}{\partial \ve{\theta}} \bigg)^{\T}\ve{R}_{\ve{\tilde{\phi}}}(\ve{\theta}) \frac{\partial \ve{\tilde{w}}(\ve{\theta})}{\partial \ve{\theta}}\notag\\
&=\bigg(\frac{\partial \ve{\mu}_{\ve{\tilde{\phi}}} (\ve{\theta}) }{\partial \ve{\theta}} \bigg)^{\T}  \frac{\partial \ve{\tilde{w}}(\ve{\theta})}{\partial \ve{\theta}}.
\end{align}
Therefore, one obtains
\begin{align}\label{exponential:replacement:wtilde}
\frac{\partial \ve{\tilde{w}}(\ve{\theta})}{\partial \ve{\theta}}=\ve{R}^{-1}_{\ve{\tilde{\phi}}}(\ve{\theta}) \frac{\partial \ve{\mu}_{\ve{\tilde{\phi}}} (\ve{\theta}) }{\partial \ve{\theta}},
\end{align}
such that the conservative information matrix is
\begin{align}\label{pessimistic:fisher:information:matrix}
\ve{\tilde{F}_u}(\ve{\theta}) = \bigg(\frac{\partial \ve{\mu}_{\ve{\tilde{\phi}}}(\ve{\theta})}{ \partial \ve{\theta}} \bigg)^{\T} \ve{R}_{\ve{\tilde{\phi}}}^{-1}(\ve{\theta}) \frac{\partial \ve{\mu}_{\ve{\tilde{\phi}}}(\ve{\theta})}{ \partial \ve{\theta}}.
\end{align}
Together with \eqref{fisher:information:inequality:exp:rep}, this results in the information inequality
\begin{align}\label{quantitative:fisher:bound}
\ve{F_u}(\ve{\theta}) &\succeq \bigg(\frac{\partial \ve{\mu}_{\ve{\tilde{\phi}}}(\ve{\theta})}{ \partial \ve{\theta}} \bigg)^{\T} \ve{R}_{\ve{\tilde{\phi}}}^{-1}(\ve{\theta}) \frac{\partial \ve{\mu}_{\ve{\tilde{\phi}}}(\ve{\theta})}{ \partial \ve{\theta}},
\end{align}
which was stated in \cite[Sec. 2]{Zografos94} without giving a derivation.

\subsection{Remarks}
As quantitative evaluation of the information matrix \eqref{pessimistic:fisher:information:matrix} exclusively requires characterization of the mean \eqref{equivalence:mean}, its derivative and the covariance matrix \eqref{equivalence:covariance}, the art of applying the Fisher information matrix lower bound \eqref{quantitative:fisher:bound} is to choose the surrogate statistics \eqref{definition:auxiliary:statistics} such that the obtained conservative approximation is analytically or computationally tractable and as tight as possible. For some insight on the latter, note that for the case that $p_{\ve{u}}(\ve{u};\ve{\theta})$ is part of the exponential family, with the result \eqref{pessimistic:fisher:information:matrix}, it holds that the exact Fisher information matrix of the original data model is, besides \eqref{fish:identity:vec:compact}, given by
\begin{align}\label{fish:identity:covariance}
\ve{F_u}(\ve{\theta}) = \bigg(\frac{\partial \ve{\mu}_{\ve{\phi}}(\ve{\theta})}{ \partial \ve{\theta}} \bigg)^{\T} \ve{R}_{\ve{\phi}}^{-1}(\ve{\theta}) \frac{\partial \ve{\mu}_{\ve{\phi}}(\ve{\theta})}{ \partial \ve{\theta}}.
\end{align}
Therefore, if the actual distribution model $p_{\ve{u}}(\ve{u};\ve{\theta})$ is part of the exponential family \eqref{exp:family:vec} and its sufficient statistics $\ve{\tilde{\phi}}(\ve{u})=\ve{\phi}(\ve{u})$ are used as surrogate statistics to construct an equivalent exponential family $\tilde{p}_{\ve{u}}(\ve{u};\ve{\theta})$ via the mean vector \eqref{equivalence:mean} and the covariance matrix \eqref{equivalence:covariance}, one obtains equality in \eqref{quantitative:fisher:bound}. Consequently, the sufficient statistics of conventional exponential family distribution models (e.g., products, absolute values, log-values) provide a guideline for the choice of the statistics \eqref{definition:auxiliary:statistics}. 

Note that the identity \eqref{fish:identity:covariance} enables accurate estimation-theoretic performance analysis within the exponential family without explicit characterization of the likelihood \eqref{exp:family:vec}. For example, the univariate Bernoulli, binomial (with known number of trails), and Gaussian (with known variance) distribution share the same sufficient statistic $\phi(u)=u$ with respect to the unknown parameter $\theta$, while their likelihood functions are distinct and defined on different supports $\mathcal{U}$. However, with an equivalent parametric mean and variance on $\tilde{\phi}(u)=\phi(u)=u$, independently of the support $\mathcal{U}$, the concept of the equivalent exponential family together with the identity \eqref{fish:identity:covariance} unifies these three distributions within one probabilistic framework which accurately characterizes the parameter-specific information contained in their samples. 

For distribution models which can not be written according to \eqref{exp:family:vec}, the statistics $\ve{\tilde{\phi}}(\ve{u})$ provide a probabilistic portrait on the good-natured canvas of the exponential family. From an engineering point of view, the free choice of the statistics $\ve{\tilde{\phi}}(\ve{u})$ forms a particular strength of the lower bound \eqref{quantitative:fisher:bound}. Suppose a scenario where the engineer wishes to evaluate parameter estimation performance as a function of specific statistics \eqref{definition:auxiliary:statistics} because they are mathematically tractable, physically interpretable, or efficiently implementable. In such a situation, the conservative information matrix \eqref{pessimistic:fisher:information:matrix} provides a way to quantitatively determine the accuracy level, which is guaranteed to be achievable through consistent estimation algorithms (see Sec. \ref{sec:achievability}). From such an application-oriented perspective, the lower bound \eqref{quantitative:fisher:bound} can be understood as determining minimum Fisher information under a constraint on the applicable statistics \eqref{definition:auxiliary:statistics}. So if the only information one has is that the engineer is restricted to the functions \eqref{definition:auxiliary:statistics} when acquiring and preprocessing the measurement data, then the matrix \eqref{pessimistic:fisher:information:matrix} is associated with the characterization of a guaranteed achievable sensitivity level when further processing the data.
\subsection{Special Cases of the Fisher Information Lower Bound}
Several special cases of the Fisher information lower bound \eqref{quantitative:fisher:bound} can be found in the literature. Considering a univariate model $p_{u}(u;\theta)$ with a single parameter $\theta\in\fieldR$, while using the statistic $\tilde{\phi}(u)=u$, yields \cite{Sankaran64,Jarrett84,Stein14}
\begin{align}\label{bound:sankaran}
F_u(\theta)\geq \frac{1}{\sigma_u^2(\theta)} \bigg(\frac{\partial {\mu_u}(\theta)}{ \partial \theta}\bigg)^{2},
\end{align}
with mean and variance
\begin{align}\label{univariate:mean}
\mu_u(\theta)&=\exdi{u;\theta}{u},\\
\label{univariate:variance}
\sigma_u^2(\theta)&=\exdi{u;\theta}{\big( u-\mu_u(\theta) \big)^2}.
\end{align}
Another compact bounding technique for the univariate-data single-parameter case, involving the skewness and kurtosis of the data-generating model $p_{u}(u;\theta)$, is discussed in \cite{Stein15}. As this lower bound features the derivative of \eqref{univariate:mean} and \eqref{univariate:variance}, it provides an approximation accuracy equivalent to the Fisher information lower bound obtained from \eqref{quantitative:fisher:bound} with the two statistics $\tilde{\phi}_1(u)=u$ and $\tilde{\phi}_2(u)=(u -\mu_{u}(\theta) )^2$.

A univariate-data single-parameter version of \eqref{quantitative:fisher:bound}, i.e.,
\begin{align}\label{bound:jarret}
F_u(\theta) \geq \bigg(\frac{\partial \ve{\mu}_{\ve{\tilde{\phi}}}(\theta)}{ \partial {\theta}} \bigg)^{\T} \ve{R}_{\ve{\tilde{\phi}}}^{-1}({\theta}) \frac{\partial \ve{\mu}_{\ve{\tilde{\phi}}}({\theta})}{ \partial {\theta}},
\end{align}
where $\ve{\tilde{\phi}}(u)$ contains the raw moment statistics $\tilde{\phi}_l(u)=u^l$ is derived in \cite{Jarrett84} by rearranging the Cram\'er-Rao inequality. Further, \cite{Jarrett84} indicates the possibility to extend \eqref{bound:jarret} to the multiple-parameter case with more general statistics, while requiring statistical orthogonality of \eqref{definition:auxiliary:statistics} (i.e., the covariance matrix \eqref{equivalence:covariance} is diagonal) and a reference statistic $\tilde{\phi}_0(u)=1, \forall u\in\mathcal{U}$.

A lower bound for probabilistic models $p_{\ve{u}}(\ve{u};\ve{\theta})$ with multivariate observations and multiple parameters
\begin{align}\label{bound:stein}
\ve{F_u}(\ve{\theta}) \succeq \bigg(\frac{\partial \ve{\mu}_{\ve{u}}(\ve{\theta})}{ \partial \ve{\theta}} \bigg)^{\T} \ve{R}_{\ve{u}}^{-1}(\ve{\theta}) \frac{\partial \ve{\mu}_{\ve{u}}(\ve{\theta})}{ \partial \ve{\theta}},
\end{align}
where
\begin{align}
\ve{\mu}_{\ve{u}}(\ve{\theta})&=\exdi{\ve{u};\ve{\theta}}{  \ve{u} },\\
\ve{R}_{\ve{u}}(\ve{\theta}) &=\exdi{\ve{u};\ve{\theta}}{\big(\ve{u}-\ve{\mu}_{\ve{u}}(\ve{\theta})\big)\big(\ve{u}-\ve{\mu}_{\ve{u}}(\ve{\theta})\big)^{\T} },
\end{align}
$\ve{\mu}_{\ve{u}}(\ve{\theta}) \in \fieldR^{M}$, $\ve{R}_{\ve{u}}(\ve{\theta})\in \fieldR^{M \times M}$, was derived in \cite{Stein14} and applied to a sensor design problem in the context of channel parameter estimation (shift in mean problem) with hard-limited radio measurements \cite{SteinWCL14}. Note that the lower bound \eqref{bound:stein} is obtained from \eqref{quantitative:fisher:bound} by using the statistics $\ve{\tilde{\phi}}(\ve{u})=\ve{u}$, such that it forms the generalization of \eqref{bound:sankaran} to multivariate observations and multiple parameters. Note that for the hard-limited observations \eqref{mapping:sign} of the multivariate zero-mean Gaussian distribution \eqref{multi:gauss:model}, one obtains $\ve{\mu_z}(\ve{\theta})=\ve{0}, \forall \ve{\theta}\in\ve{\Theta}$, such that for shift in covariance problems the right-hand side of the matrix inequality \eqref{bound:stein} does not provide a useful approximation. 
\subsection{Interpretation as a Gaussian Modeling Framework}
In \cite{Stein14}, it was observed that the right-hand side of the information bound \eqref{bound:stein} can be interpreted as the Fisher matrix of a particular Gaussian modeling framework. Similarly, here the matrix
\begin{align}\label{pessimistic:fisher:matrix:gauss:interpretation}
\ve{\tilde{F}_u}(\ve{\theta}) = \bigg(\frac{\partial \ve{\mu}_{\ve{\tilde{\phi}}}(\ve{\theta})}{ \partial \ve{\theta}} \bigg)^{\T} \ve{R}_{\ve{\tilde{\phi}}}^{-1}(\ve{\theta}) \frac{\partial \ve{\mu}_{\ve{\tilde{\phi}}}(\ve{\theta})}{ \partial \ve{\theta}}
\end{align}
is the Fisher information matrix of an $\tilde{L}$-variate vector $\ve{\tilde{\phi}}$ following the multivariate Gaussian distribution
\begin{align}\label{gauss:model:interpretation}
p_{\ve{\tilde{\phi}}}(\ve{\tilde{\phi}};\ve{\theta})=\frac{\exp{- \frac{1}{2} (\ve{\tilde{\phi}}- \ve{\mu}_{\ve{\tilde{\phi}}}(\ve{\theta}))^{\T} \ve{R}^{-1}_{\ve{\tilde{\phi}}}(\ve{\theta})(\ve{\tilde{\phi}}- \ve{\mu}_{\ve{\tilde{\phi}}}(\ve{\theta}))}}{(2\pi)^{\frac{\tilde{L}}{2}} \sqrt{\det \big(\ve{R}_{\ve{\tilde{\phi}}}(\ve{\theta})\big) }} 
\end{align}
with the restriction
\begin{align}\label{gauss:model:interpretation:condition}
\frac{\partial \ve{R}_{\ve{\tilde{\phi}}}(\ve{\theta})}{\partial \theta_d}=\ve{0},\quad d=1,2,\ldots,D.
\end{align}
Therefore, in an asymptotic sense, an equivalent exponential family $\tilde{p}_{\ve{u}}(\ve{u};\ve{\theta})$ with sufficient statistics $\ve{\tilde{\phi}}(\ve{u})$ can be interpreted as a probabilistic portrait of the $M$-variate data model $p_{\ve{u}}(\ve{u};\ve{\theta})$ on the canvas of the $\tilde{L}$-variate Gaussian framework \eqref{gauss:model:interpretation} with the special property \eqref{gauss:model:interpretation:condition}. 
\section{Achievability of the Information Bound}\label{sec:achievability}
In theory, the Fisher information lower bound \eqref{quantitative:fisher:bound} enables conservatively analyzing the parameter-specific information contained in samples from unknown probabilistic data models. An advantage of the conservative information matrix \eqref{pessimistic:fisher:information:matrix} is that its quantitative evaluation does not require an explicit characterization of the distribution function and can be conducted just with the mean \eqref{equivalence:mean} and the covariance \eqref{equivalence:covariance} at hand. In practice, however, also the question arises as to how efficient parameter estimation should be carried out without explicit knowledge of the likelihood. 

To this end, it is assumed that a dataset with $N$ samples
\begin{align}\label{data:observations:matrix}
\ve{U}=\begin{bmatrix} \ve{u}_1 &\ve{u}_2 &\ldots &\ve{u}_N \end{bmatrix},
\end{align}
each independently drawn from the $M$-variate distribution function $p_{\ve{u}}(\ve{u};\ve{\theta}_t)$ with unknown true parameters $\ve{\theta}_t$, is available. Further, the statistics \eqref{definition:auxiliary:statistics} have been chosen such that the mean \eqref{equivalence:mean} and covariance matrix \eqref{equivalence:covariance} are available in parametric form. As a preprocessing step, data compression is performed by computing the empirical mean of the dataset
\begin{align}\label{data:compressed}
\ve{\hat{\mu}_{\ve{\tilde{\phi}}}}(\ve{U})=\frac{1}{N}\sum_{n=1}^{N} \ve{\tilde{\phi}}(\ve{u}_n)
\end{align}
and discarding the measurements \eqref{data:observations:matrix}. This reduces the dimensionality of the stored data by $\frac{\tilde{L}}{MN}$. 

The algorithm for estimating the distribution model parameters $\ve{\theta}$ is designed according to the principle of maximum-likelihood where the equivalent exponential family distribution is used in the sense of a quasi-likelihood \cite{Wedderburn74}, i.e., one solves
\begin{align}\label{definition:cmle}
\ve{\hat{\theta}}(\ve{U})
&=\arg\max_{\ve{\theta} \in \ve{\Theta}} \ln \tilde{p}_{\ve{U}}(\ve{U};\ve{\theta})\notag\\
&=\arg\max_{\ve{\theta} \in \ve{\Theta}} \sum_{n=1}^{N}\ln \tilde{p}_{\ve{u}}(\ve{u}_n;\ve{\theta}).
\end{align}
The solution $\ve{\theta}^{\star}$ of this optimization problem satisfies
\begin{align}\label{cmle:solution:zero:root}
\frac{ \partial \ln \tilde{p}_{\ve{U}}(\ve{U};\ve{\theta}^{\star})}{\partial \ve{\theta}}=\sum_{n=1}^{N} \frac{ \partial \ln \tilde{p}_{\ve{u}}(\ve{u}_n;\ve{\theta}^{\star})}{\partial \ve{\theta}}=\ve{0}^{\T},
\end{align}
where, with \eqref{exponential:replacement:score:wtilde}, \eqref{exponential:replacement:wtilde}, and \eqref{data:compressed}
\begin{align}
&\sum_{n=1}^{N} \bigg(\frac{ \partial \ln \tilde{p}_{\ve{u}}(\ve{u}_n;\ve{\theta})}{\partial \ve{\theta}}\bigg)^{\T}=\notag\\
&=\sum_{n=1}^{N} \bigg(\frac{\partial \ve{\mu}_{\ve{\tilde{\phi}}}(\ve{\theta})}{ \partial \ve{\theta}} \bigg)^{\T} \ve{R}_{\ve{\tilde{\phi}}}^{-1}(\ve{\theta})\big(\ve{\tilde{\phi}}(\ve{u}_n) - \ve{\mu}_{\ve{\tilde{\phi}}}(\ve{\theta})  \big)\notag\\
& =N \bigg(\frac{\partial \ve{\mu}_{\ve{\tilde{\phi}}}(\ve{\theta})}{ \partial \ve{\theta}} \bigg)^{\T} \ve{R}_{\ve{\tilde{\phi}}}^{-1}(\ve{\theta}) \big(\ve{\hat{\mu}_{\ve{\tilde{\phi}}}}(\ve{U}) - \ve{\mu}_{\ve{\tilde{\phi}}}(\ve{\theta})  \big).
\end{align}
It is, therefore, evident that the solution for \eqref{definition:cmle} can be found exclusively on the basis of the mean \eqref{equivalence:mean} and the covariance \eqref{equivalence:covariance} while explicit knowledge of the likelihood is not required.
\subsection{Consistency}
By the law of large numbers
\begin{align}\label{score:convergence}
\frac{1}{N}\sum_{n=1}^{N} \frac{ \partial \ln \tilde{p}_{\ve{u}}(\ve{u}_n;\ve{\theta})}{\partial \ve{\theta}} &\overset{a.s.}{\to} \exdi{\ve{u};\ve{\theta}_t}{ \frac{ \partial \ln \tilde{p}_{\ve{u}}(\ve{u};\ve{\theta}) }{\partial \ve{\theta}}  },
\end{align}
where $\overset{a.s.}{\to}$ denotes almost sure convergence. Therefore,
\begin{align}
&\bigg(\frac{\partial \ve{\mu}_{\ve{\tilde{\phi}}}(\ve{\theta})}{ \partial \ve{\theta}} \bigg)^{\T} \ve{R}_{\ve{\tilde{\phi}}}^{-1}(\ve{\theta}) \big(\ve{\hat{\mu}_{\ve{\tilde{\phi}}}}(\ve{U}) - \ve{\mu}_{\ve{\tilde{\phi}}}(\ve{\theta})  \big)\overset{a.s.}{\to} \notag\\
&\bigg(\frac{\partial \ve{\mu}_{\ve{\tilde{\phi}}}(\ve{\theta})}{ \partial \ve{\theta}} \bigg)^{\T} \ve{R}_{\ve{\tilde{\phi}}}^{-1}(\ve{\theta}) \big(\ve{\mu}_{\ve{\tilde{\phi}}}(\ve{\theta}_t) - \ve{\mu}_{\ve{\tilde{\phi}}}(\ve{\theta})  \big),
\end{align}
such that for the root in \eqref{cmle:solution:zero:root}
\begin{align}\label{cmle:solution:consistent}
\ve{\theta}^{\star} \overset{a.s.}{\to} \ve{\theta}_t. 
\end{align}
This shows that solving \eqref{definition:cmle} provides a consistent estimator.
\subsection{Efficiency}
For the analysis of the error covariance of the estimator \eqref{definition:cmle}, with the notational convention 
\begin{align}
\frac{\partial \ve{f}(\ve{x}')}{\partial \ve{x}}= \left. \frac{\partial \ve{f}(\ve{x})}{\partial \ve{x}} \right|_{\ve{x}=\ve{x}'},
\end{align}
the Taylor expansion around the true parameter $\ve{\theta}_t$ is
\begin{align}
\frac{\partial \ln \tilde{p}_{\ve{U}}(\ve{U};\ve{\theta})}{\partial \ve{\theta}}  &= \frac{\partial \ln \tilde{p}_{\ve{U}}(\ve{U};\ve{\theta}_t)}{\partial \ve{\theta}} + (\ve{\theta} - \ve{\theta}_t)^{\T} \frac{\partial^2 \ln \tilde{p}_{\ve{U}}(\ve{U};\ve{\breve{\theta}})}{\partial \ve{\theta}^2},
\end{align}
where $\ve{\breve{\theta}}$ lies on the line between $\ve{\theta}$ and $\ve{\theta}_t$. Due to \eqref{cmle:solution:zero:root},
\begin{align}
\frac{\partial \ln \tilde{p}_{\ve{U}}(\ve{U};\ve{\theta}_t)}{\partial \ve{\theta}} =-(\ve{\theta}^{\star} - \ve{\theta}_t)^{\T} \frac{\partial^2 \ln \tilde{p}_{\ve{U}}(\ve{U};\ve{\breve{\theta}})}{\partial \ve{\theta}^2} ,
\end{align}
such that
\begin{align}
&\sqrt{N}(\ve{\theta}^{\star} - \ve{\theta}_t)^{\T}=\notag\\
&= \Bigg(\frac{1}{\sqrt{N}} \frac{\partial \ln \tilde{p}_{\ve{U}}(\ve{U};\ve{\theta}_t)}{\partial \ve{\theta}}\Bigg) \Bigg(-\frac{1}{N} \frac{\partial^2 \ln \tilde{p}_{\ve{U}}(\ve{U};\ve{\breve{\theta}})}{\partial \ve{\theta}^2} \Bigg)^{-1}.
\end{align}
With the law of large numbers, consistency \eqref{cmle:solution:consistent}, and \eqref{fish:expo:info:ident2},
\begin{align}
&-\frac{1}{N}\frac{\partial^2 \ln \tilde{p}_{\ve{U}}(\ve{U};\ve{\breve{\theta}})}{\partial \ve{\theta}^2} =-\frac{1}{N} \sum_{n=1}^{N} \frac{\partial^2 \ln \tilde{p}_{\ve{u}}(\ve{u}_n;\ve{\breve{\theta}})}{\partial \ve{\theta}^2} 
\end{align}
converges to the conservative information matrix
\begin{align}
\ve{\tilde{F}_u}(\ve{\theta}_t) = - \exdi{\ve{u};\ve{\theta}_t}{\frac{\partial^2 \ln \tilde{p}_{\ve{u}}(\ve{u};\ve{\theta}_t)}{\partial \ve{\theta}^2} }.
\end{align}
Due to the central limit theorem,
\begin{align}
\frac{1}{\sqrt{N}} \frac{\partial \ln \tilde{p}_{\ve{U}}(\ve{U};\ve{\theta}_t)}{\partial \ve{\theta}} =  \frac{1}{\sqrt{N}}  \sum_{n=1}^{N} \frac{\partial \ln \tilde{p}_{\ve{u}}(\ve{u}_n;\ve{\theta}_t)}{\partial \ve{\theta}}
\end{align}
converges to a multivariate Gaussian random variable with zero mean (due to regularity) and covariance
\begin{align}
&\frac{1}{N}\exdi{\ve{U};\ve{\theta}_t}{ \bigg( \sum_{n=1}^{N}\frac{\partial \ln \tilde{p}_{\ve{u}}(\ve{u}_n;\ve{\theta}_t)}{\partial \ve{\theta}}  \bigg)^{\T}  \sum_{n=1}^{N} \frac{\partial \ln \tilde{p}_{\ve{u}}(\ve{u}_n;\ve{\theta}_t)}{\partial \ve{\theta}}  }=\notag\\
&=\exdi{\ve{u};\ve{\theta}_t}{  \bigg(  \frac{\partial \ln \tilde{p}_{\ve{u}}(\ve{u};\ve{\theta}_t)}{\partial \ve{\theta}} \bigg)^{\T}  \frac{\partial \ln \tilde{p}_{\ve{u}}(\ve{u};\ve{\theta}_t)}{\partial \ve{\theta}} }\notag\\
&=\ve{\tilde{F}_u}(\ve{\theta}_t),
\end{align}
where the last step is due to \eqref{fish:expo:info:ident2}.
Therefore, with Slutsky's theorem and $\overset{d.}{\sim}$ denoting convergence in distribution, it follows that
\begin{align}
&\sqrt{N}(\ve{\theta}^{\star} - \ve{\theta}_t)\overset{d.}{\sim}\mathcal{N}\big(\ve{0},\ve{\tilde{F}}_{\ve{u}}^{-1}(\ve{\theta}_t)\big).
\end{align}
This shows that the solution of \eqref{definition:cmle} is asymptotically efficient regarding the conservative information matrix \eqref{pessimistic:fisher:information:matrix}.
\subsection{Connection to the Generalized Method of Moments}
Taking the inner product of \eqref{cmle:solution:zero:root}, potential solutions of the optimization problem \eqref{definition:cmle} can be identified by solving
\begin{align}\label{cmle:root:minimization}
&\ve{\hat{\theta}}(\ve{U})=\notag\\
&=\arg \min_{\ve{\theta}\in\ve{\Theta}}\big(\ve{\hat{\mu}_{\ve{\tilde{\phi}}}}(\ve{U}) - \ve{\mu}_{\ve{\tilde{\phi}}}(\ve{\theta})\big)^{\T}\ve{\Pi}(\ve{\theta})\big(\ve{\hat{\mu}_{\ve{\tilde{\phi}}}}(\ve{U}) - \ve{\mu}_{\ve{\tilde{\phi}}}(\ve{\theta})\big)
\end{align}
with
\begin{align}
\ve{\Pi}(\ve{\theta})&=\ve{R}_{\ve{\tilde{\phi}}}^{-1}(\ve{\theta}) \frac{\partial \ve{\mu}_{\ve{\tilde{\phi}}}(\ve{\theta})}{ \partial \ve{\theta}} \bigg( \frac{\partial \ve{\mu}_{\ve{\tilde{\phi}}}(\ve{\theta})}{ \partial \ve{\theta}} \bigg)^{\T} \ve{R}_{\ve{\tilde{\phi}}}^{-1}(\ve{\theta}).
\end{align}
This expression is equivalent to Hansen's estimator \cite{Hansen82}
\begin{align}\label{estimator:hansen}
&\ve{\hat{\theta}}(\ve{U})=\notag\\
&=\arg \min_{\ve{\theta}\in\ve{\Theta}} \bigg(\frac{1}{N}\sum_{n=1}^{N} \ve{f}(\ve{u}_n;\ve{\theta})\bigg)^{\T}\ve{H}(\ve{\theta}) \bigg(\frac{1}{N}\sum_{n=1}^{N} \ve{f}(\ve{u}_n;\ve{\theta})\bigg)
\end{align}
when using
\begin{align}
\ve{f}(\ve{u};\ve{\theta})&=\ve{\tilde{\phi}}(\ve{u})-\ve{\mu}_{\ve{\tilde{\phi}}}(\ve{\theta}),\\
\ve{H}(\ve{\theta}) &= \ve{\Pi}(\ve{\theta}).
\end{align}
Note that the estimator \eqref{estimator:hansen} is designed with the orthogonality condition
\begin{align}\label{condition:orthogonality}
\exdi{\ve{u};\ve{\theta}_t}{\ve{f}(\ve{u};\ve{\theta}_t)}=\ve{0}
\end{align}
and generalizes Pearson's estimation method \cite{Pearson94} from raw moments to arbitrary statistics. With \eqref{cmle:root:minimization} it can be seen that Hansen's estimator \eqref{estimator:hansen} alternatively follows from approximating the data-generating model $p_{\ve{u}}(\ve{u};\ve{\theta})$ by an equivalent exponential family $\tilde{p}_{u}(\ve{u};\ve{\theta})$ with sufficient statistics $\ve{\tilde{\phi}}(\ve{u})$ and subsequently following the principle of maximizing the likelihood. Therefore, the equivalent exponential family $\tilde{p}_{u}(\ve{u};\ve{\theta})$ forms the unifying link between Pearson's method of moments and Fisher's competing concept of maximum-likelihood. This explicit observation supports a claim that was made in the context of maximum-entropy techniques \cite[p. 141]{Kapur92}.
\section{Conservative Information Matrix for Hard-limited Multivariate Gaussian Data}\label{sec:gaussian:data}
In the following, the computation of a conservative Fisher information matrix \eqref{pessimistic:fisher:information:matrix} for the case of hard-limited zero-mean multivariate Gaussian data \eqref{mapping:sign} is discussed. In particular, a specific choice of the auxiliary statistics \eqref{aux:statistics:quantizer}, the calculation of the required mean \eqref{equivalence:mean} and covariance matrix \eqref{equivalence:covariance}, and an approximation that can be used to accelerate the evaluation are outlined.
\subsection{Statistics for Hard-limited Gaussian Data}
The zero-mean Gaussian model \eqref{multi:gauss:model} is part of the exponential family \eqref{exp:family:vec}, while its sufficient statistics \eqref{sufficient:stat:gaussian:zero:mean} consist of the pairwise products between the $M$ random variables. Multivariate binary distributions also admit a likelihood function with the structure \eqref{exp:family:vec}. However, the sufficient statistics $\ve{\phi}(\ve{z})$, besides the pairwise products, also contain all possible higher-order products between the $M$ random variables, see, e.g., \cite{Dai13}. As such, the number of sufficient statistics $L$ for a multivariate binary distribution, in general, grows according to $\mathcal{O}(2^M)$. To control the statistical complexity $L$, here a zero-mean quadratic exponential family \cite{Cox94} is used as a surrogate distribution model. Therefore, the statistics \eqref{definition:auxiliary:statistics} are restricted to the pairwise products, such that $\tilde{L}$ only grows with $\mathcal{O}(M^2)$. 

In detail, this means that an equivalent exponential family with the $\tilde{L}=\frac{M}{2}(M-1)$ statistics
\begin{align}\label{aux:statistics:quantizer}
\ve{\tilde{\phi}}(\ve{z})&=\ve{\Phi}\vec{\ve{z} \ve{z}^{\T}}
\end{align}
is employed, where $\ve{\Phi}\in\{0, 1\}^{\tilde{L} \times M^2}$ denotes an elimination matrix discarding the $\frac{M}{2}(M-1)$ duplicate and the $M$ constant diagonal entries of the random variables' outer product $\ve{z}\ve{z}^{\T}$.

\subsection{Parametric Mean and Covariance of Quadratic Statistics}
With \eqref{aux:statistics:quantizer}, the mean \eqref{equivalence:mean} of the statistics \eqref{definition:auxiliary:statistics} is
\begin{align}
\ve{\mu}_{\ve{\tilde{\phi}}}(\ve{\theta})&=\exdi{\ve{z};\ve{\theta}}{\ve{\tilde{\phi}}(\ve{z})} = \ve{\Phi}\vec{\ve{R}_{\ve{z}}(\ve{\theta})},
\end{align}
where the covariance matrix of the quantized zero-mean data is defined as
\begin{align}
\ve{R}_{\ve{z}}(\ve{\theta})=\exdi{\ve{z};\ve{\theta}}{\ve{z}\ve{z}^{\T}}.
\end{align}
Through the arcsine law \cite[pp. 284]{Thomas69}, the quantized covariance can be calculated from the unquantized covariance \eqref{definition:gauss:covariance:matrix}
\begin{align}\label{covariance:quantized}
\ve{R}_{\ve{z}}(\ve{\theta})=\frac{2}{\pi} \arcsin{ \ve{C}_{\ve{y}}(\ve{\theta}) },
\end{align}
with the correlation matrix (normalized covariance matrix)
\begin{align}\label{correlation:matrix}
\ve{C}_{\ve{y}}(\ve{\theta})=\diag{\ve{R}_{\ve{y}}(\ve{\theta})}^{-\frac{1}{2}} \ve{R}_{\ve{y}}(\ve{\theta}) \diag{\ve{R}_{\ve{y}}(\ve{\theta})}^{-\frac{1}{2}}.
\end{align}
Here, for a square matrix $\ve{B}$, the operator $\diag{\ve{B}}$ provides a matrix $\ve{B}'$ with diagonal elements equal to the ones of the matrix $\ve{B}$ and all the off-diagonal elements equal to zero. 

For the derivative of the statistics' mean
\begin{align}
\frac{\partial \ve{\mu}_{\ve{\tilde{\phi}}}(\ve{\theta})}{\partial \ve{\theta}} =\ve{\Phi} \frac{\partial \vec{ \ve{R}_{\ve{z}}(\ve{\theta}) }}{\partial \ve{\theta}} ,
\end{align}
the individual columns $d=1,\ldots,D$ are
\begin{align}
\left[ \frac{\partial \ve{\mu}_{\ve{\tilde{\phi}}}(\ve{\theta})}{\partial \ve{\theta}} \right]_{d} &= \frac{\partial \ve{\mu}_{\ve{\tilde{\phi}}}(\ve{\theta})}{\partial \theta_d} = \ve{\Phi}\vec{ \frac{\partial \ve{R}_{\ve{z}}(\ve{\theta}) }{\partial \theta_d}}.
\end{align}
The off-diagonal entries of the derivative of \eqref{covariance:quantized} are 
\begin{align}
&\left[ \frac{\partial \ve{R}_{\ve{z}}(\ve{\theta})}{\partial \theta_d} \right]_{ij} =  \frac{2}{\pi} \frac{\left[ \frac{\partial \ve{C}_{\ve{y}}(\ve{\theta})}{\partial \theta_d} \right]_{ij}}{\sqrt{1-  \left[\ve{C}_{\ve{y}}(\ve{\theta})\right]_{ij}^2}}, \quad i \neq j,
\end{align}
while the diagonal entries ($i=j$) are equal to zero. For the covariance \eqref{equivalence:covariance}, it is required to compute
\begin{align}\label{uncentered:covariance:statistics:gauss:hardlim}
\exdi{\ve{z};\ve{\theta}}{\ve{\tilde{\phi}}(\ve{z}) \ve{\tilde{\phi}}^{\T}(\ve{z})}=\ve{\Phi}\exdi{\ve{z};\ve{\theta}}{\vec{\ve{z} \ve{z}^{\T}}\vec{\ve{z} \ve{z}^{\T}}^{\T}} \ve{\Phi}^{\T},
\end{align}
which implies to evaluate quadrivariate expected values of the form
\begin{align}\label{binary:quadrivariate:mean}
\exdi{\ve{z};\ve{\theta}}{z_i z_j z_k z_q},\quad\quad i, j, k, q \in \{1,\ldots, M\}.
\end{align}
For the cases $i=j=k=q$ and $i=j \neq k=q$, one obtains
\begin{align}
\exdi{\ve{z};\ve{\theta}}{z_i z_j z_k z_q}&=1.
\end{align}
If $i=j=k \neq q$ or $i=j\neq k \neq q$, the arcsine law provides
\begin{align}
\exdi{\ve{z};\ve{\theta}}{z_i z_j z_k z_q}
&=\frac{2}{\pi} \arcsin{\left[\ve{C}_{\ve{y}}(\ve{\theta})\right]_{kq}}.
\end{align}
The cases  $i \neq j \neq k \neq q$ require to calculate
\begin{align}
\exdi{\ve{z};\ve{\theta}}{z_i z_j z_k z_q}&=\Prob{z_i z_j z_k z_q=1}\notag\\
&-\Prob{z_i z_j z_k z_q=-1},
\end{align}
which involves the evaluation of the $2^4=16$ orthant probabilities of a zero-mean quadrivariate Gaussian variable
\begin{align}
\ve{y}'=\begin{bmatrix} y_i &y_j &y_k &y_q \end{bmatrix}^{\T},
\end{align}
with covariance matrix
\begin{align}
&\ve{R}_{\ve{y}'}(\ve{\theta})=\notag\\
&=\begin{bmatrix} 
\left[\ve{R}_{\ve{y}}(\ve{\theta})\right]_{ii} &\left[\ve{R}_{\ve{y}}(\ve{\theta})\right]_{ij} &\left[\ve{R}_{\ve{y}}(\ve{\theta})\right]_{ik} &\left[\ve{R}_{\ve{y}}(\ve{\theta})\right]_{iq}\\
\left[\ve{R}_{\ve{y}}(\ve{\theta})\right]_{ji} &\left[\ve{R}_{\ve{y}}(\ve{\theta})\right]_{jj} &\left[\ve{R}_{\ve{y}}(\ve{\theta})\right]_{jk} &\left[\ve{R}_{\ve{y}}(\ve{\theta})\right]_{jq}\\
\left[\ve{R}_{\ve{y}}(\ve{\theta})\right]_{ki} &\left[\ve{R}_{\ve{y}}(\ve{\theta})\right]_{kj} &\left[\ve{R}_{\ve{y}}(\ve{\theta})\right]_{kk} &\left[\ve{R}_{\ve{y}}(\ve{\theta})\right]_{kq}\\
\left[\ve{R}_{\ve{y}}(\ve{\theta})\right]_{qi} &\left[\ve{R}_{\ve{y}}(\ve{\theta})\right]_{qj}&\left[\ve{R}_{\ve{y}}(\ve{\theta})\right]_{qk} &\left[\ve{R}_{\ve{y}}(\ve{\theta})\right]_{qq}\\
\end{bmatrix}.
\end{align}
A mathematical expression for the orthant probabilities of the zero-mean quadrivariate Gaussian distribution model 
\begin{align}
\ve{y}'\sim\distN(\ve{0},\ve{R}_{\ve{y}'}(\ve{\theta})), 
\end{align}
consisting of four one-dimensional integrals, is given in \cite{Sinn11}. 

\subsection{Heuristic Approximation for Hard-limited Gaussian Distributions}
As with the statistics \eqref{aux:statistics:quantizer} the size of the covariance matrix \eqref{equivalence:covariance} grows quadratically with the number of variables $M$, while for \eqref{pessimistic:fisher:information:matrix} its inverse is required, a heuristic approximation is discussed which simplifies computation of the information matrix \eqref{pessimistic:fisher:information:matrix} when considering hard-limited observations \eqref{mapping:sign} of the multivariate zero-mean Gaussian model \eqref{multi:gauss:model}. The approximation is based on the observation that the entries of the zero-mean Gaussian Fisher information matrix are
\begin{align}
&\left[ \ve{F_y}(\ve{\theta}) \right]_{ij}=\notag\\
&=\frac{1}{2} \vec{ \frac{\partial \ve{R}_{\ve{y}}(\ve{\theta})}{\partial \theta_i} }^{\T} \vec{ \ve{R}_{\ve{y}}^{-1}(\ve{\theta}) \frac{\partial \ve{R}_{\ve{y}}(\ve{\theta})}{\partial \theta_j} \ve{R}_{\ve{y}}^{-1}(\ve{\theta})}\notag\\
&=\frac{1}{2} \vec{ \frac{\partial \ve{R}_{\ve{y}}(\ve{\theta})}{\partial \theta_i} }^{\T}  \big(\ve{R}_{\ve{y}}^{-1}(\ve{\theta}) \otimes \ve{R}_{\ve{y}}^{-1}(\ve{\theta}) \big) \vec{ \frac{\partial \ve{R}_{\ve{y}}(\ve{\theta})}{\partial \theta_j} }.
\end{align}
Therefore, with the notational convention
\begin{align}
\ve{r}_{\ve{y}}(\ve{\theta})=\vec{\ve{R}_{\ve{y}}(\ve{\theta})},
\end{align}
the Fisher information matrix with unquantized Gaussian measurements is characterized by
\begin{align}\label{fisher:matrix:gaussian:full:compact}
\ve{F_y}(\ve{\theta}) =\frac{1}{2}  \bigg( \frac{\partial \ve{r}_{\ve{y}}(\ve{\theta})}{\partial \ve{\theta}} \bigg)^{\T}  \big(\ve{R}_{\ve{y}}(\ve{\theta}) \otimes \ve{R}_{\ve{y}}(\ve{\theta}) \big)^{-1} \frac{\partial \ve{r}_{\ve{y}}(\ve{\theta})}{\partial \ve{\theta}}.
\end{align}
Comparing with \eqref{fish:identity:covariance}, this shows that for the multivariate Gaussian distribution \eqref{multi:gauss:model}, with its sufficient statistics \eqref{sufficient:stat:gaussian:zero:mean}, one obtains
\begin{align}
\ve{R}_{\ve{\phi}}(\ve{\theta})&= 2 \big(\ve{R}_{\ve{y}}(\ve{\theta}) \otimes \ve{R}_{\ve{y}}(\ve{\theta}) \big).
\end{align}
Such a structure of the covariance matrix \eqref{equivalence:covariance} is advantageous as, due to the properties of the Kronecker product $\otimes$, it decomposes inversion of $\ve{R}_{\ve{\phi}}(\ve{\theta})$ into the inversion of a smaller matrix, i.e.,
\begin{align}
\ve{R}_{\ve{\phi}}^{-1}(\ve{\theta})&= \frac{1}{2} \big(\ve{R}_{\ve{y}}^{-1}(\ve{\theta}) \otimes \ve{R}_{\ve{y}}^{-1}(\ve{\theta}) \big).
\end{align}
For hard-limited zero-mean Gaussian observations \eqref{mapping:sign}, this motivates using the $\tilde{L}=M^2$ statistics
\begin{align}
\ve{\tilde{\phi}}(\ve{z})&=\vec{\ve{z} \ve{z}^{\T}},
\end{align}
and approximating the resulting covariance matrix by
\begin{align}\label{approximation:quantized:covariance:aux:statistics}
\ve{R}_{\ve{\tilde{\phi}}}(\ve{\theta})&\approx 2 \big(\ve{R}_{\ve{z}}(\ve{\theta}) \otimes \ve{R}_{\ve{z}}(\ve{\theta}) \big),
\end{align}
such that
\begin{align}\label{approximate:fisher:hardlimited}
\ve{\tilde{F}_z}(\ve{\theta}) \approx \frac{1}{2}  \bigg( \frac{\partial \ve{r}_{\ve{z}}(\ve{\theta})}{\partial \ve{\theta}} \bigg)^{\T}  \big(\ve{R}_{\ve{z}}^{-1}(\ve{\theta}) \otimes \ve{R}_{\ve{z}}^{-1}(\ve{\theta}) \big) \frac{\partial \ve{r}_{\ve{z}}(\ve{\theta})}{\partial \ve{\theta}}
\end{align}
where
\begin{align}
\ve{r}_{\ve{z}}(\ve{\theta})=\vec{\ve{R}_{\ve{z}}(\ve{\theta})}.
\end{align}
Note that this approximation of the conservative information matrix $\ve{\tilde{F}_z}(\ve{\theta})$ is not guaranteed to preserve the estimation-theoretic inequality \eqref{fisher:information:inequality:exp:rep}, i.e., the approach can lead to an information matrix, which overestimates the inference capabilities $\ve{F_z}(\ve{\theta})$ with the data-generating model $p_{\ve{z}}(\ve{z};\ve{\theta})$. The advantage of the heuristic approximation \eqref{approximate:fisher:hardlimited}, however, is that inversion of the large covariance matrix $\ve{R}_{\ve{\tilde{\phi}}}(\ve{\theta})$ with the expected values of quadrivariate products \eqref{binary:quadrivariate:mean} is avoided by employing the approximation \eqref{approximation:quantized:covariance:aux:statistics}, using the arcsine law \eqref{covariance:quantized} followed by inversion of the matrix $\ve{R}_{\ve{z}}(\ve{\theta})$.

\section{Sensitivity Analysis for Binary Sampling}\label{sec:applications}
In the following, the application of the conservative information matrix \eqref{pessimistic:fisher:information:matrix} is demonstrated in the context of electrical engineering by assessing the achievable estimation accuracy with measurements acquired with binary sensor systems. First, a situation where the mean \eqref{equivalence:mean} and covariance \eqref{equivalence:covariance} of the statistics \eqref{definition:auxiliary:statistics} can be derived mathematically is considered. Then, a scenario is discussed where mean and covariance of user-defined statistics are obtained from calibrated measurement data\footnote{A measurement dataset is here considered calibrated if during its acquisition it is ensured that the parameters $\ve{\theta}$ of the data-generating system $p_{\ve{u}}(\ve{u};\ve{\theta})$ are known.}.

\subsection{Array Processing with Low-Complexity Binary Sensors}\label{section:binary:array:processing}
The first application example is the design of a large-scale array system with binary sensing elements for the localization of a random signal source. The processing task is formulated as a DOA parameter estimation problem \cite{MUSIC,ESPRIT,Haardt95}, where the goal is to determine the angle under which a narrow-band source signal with unknown temporal structure impinges onto a uniform linear array (ULA). The digital processing unit has only access to the sign of the analog sensor outputs \cite{BarShalom02}. Such a technical setup arises when using low-complexity $1$-bit A/D conversion at the sensors to minimize energy consumption and hardware cost of the array elements, or when hard-limiting high-resolution sensor data to reduce bandwidth for data transmission from the sensors to the central processing unit \cite{Ribeiro06}. The conservative information matrix \eqref{pessimistic:fisher:information:matrix} is used to determine the number of binary sensors required to achieve a certain parameter-specific sensitivity level characterized by an ideal reference system with $\infty$-bit A/D conversion.

For a ULA with $K\in\fieldN$ homodyne receivers, placed at a distance of half the carrier wavelength, and a single narrow-band random signal source with independent in-phase and quadrature component, the covariance of the real-valued received signals prior to hard-limiting \eqref{mapping:sign} can be modeled \cite{SteinWSA16}
\begin{align}\label{receive:covariance:doa}
\ve{R}_{\ve{y}}(\ve{\theta})&=\gamma\ve{A}(\zeta)\ve{A}^{\T}(\zeta)+\ve{I},\\
\ve{\theta}&=\begin{bmatrix}\gamma &\zeta \end{bmatrix}^{\T},
\end{align}
with $\gamma, \zeta \in\fieldR$ and $\gamma>0, \zeta \in \left[-\pi/2; \pi/2\right]$. Hereby, $\gamma$ represents the SNR and $\zeta$ the DOA parameter. The array steering matrix features an in-phase and quadrature component
\begin{align}
\ve{A}(\zeta)=\begin{bmatrix}\ve{A}^{\T}_\text{I}(\zeta) &\ve{A}^{\T}_\text{Q}(\zeta) \end{bmatrix}^{\T},
\end{align}
$\ve{A}(\zeta)\in\fieldR^{2K\times 2}$, where the sub-matrices have the form
\begin{align}
\ve{A}_\text{I}(\zeta)=\begin{bmatrix}
\cos{\nu_1(\zeta)} &\sin{\nu_1(\zeta)}\\ 
\vdots &\vdots\\ 
\cos{\nu_K(\zeta)} &\sin{\nu_K(\zeta)}
\end{bmatrix},
\end{align}
$\ve{A}_\text{I}(\zeta)\in\fieldR^{K \times 2}$, and
\begin{align}
\ve{A}_\text{Q}(\zeta)=\begin{bmatrix}
-\sin{\nu_1(\zeta)} &\cos{\nu_1(\zeta)}\\ 
\vdots &\vdots\\ 
-\sin{\nu_K(\zeta)} &\cos{\nu_K(\zeta)}
\end{bmatrix},
\end{align}
$\ve{A}_\text{Q}(\zeta)\in\fieldR^{K \times 2}$, with
\begin{align}
\nu_k(\zeta)=(k-1)\pi\sin{(\zeta)}.
\end{align}
The identity matrix in the covariance model \eqref{receive:covariance:doa} is due to the additive and spatially white measurement noise at the sensors. Note that, by using a real-valued signal model instead of a complex-valued one, the dimension of the digital array measurement data is $M=2K$.

With \eqref{receive:covariance:doa}, the received data correlation matrix 
\begin{align}
\ve{C}_{\ve{y}}(\ve{\theta})= \frac{1}{\gamma+1} \Big( \gamma\ve{A}(\zeta)\ve{A}^{\T}(\zeta)+\ve{I} \Big)
\end{align}
has the derivatives
\begin{align}
\frac{\partial \ve{C}_{\ve{y}}(\ve{\theta})}{ \partial \gamma} &=\frac{1}{(\gamma+1)^2} \Big(\ve{A}(\zeta)\ve{A}^{\T}(\zeta)-\ve{I} \Big),\\
\frac{\partial \ve{C}_{\ve{y}}(\ve{\theta})}{ \partial \zeta} &= \frac{\gamma}{\gamma+1} \bigg( \frac{\partial\ve{A}(\zeta)}{\partial \zeta}\ve{A}^{\T}(\zeta)+ \ve{A}(\zeta) \frac{\partial \ve{A}^{\T}(\zeta)}{ \partial \zeta} \bigg).
\end{align}

With the conservative information matrix \eqref{pessimistic:fisher:information:matrix} computed based on the statistics \eqref{aux:statistics:quantizer}, one can quantitatively evaluate the relative hard-limiting loss concerning both parameters by the measures
\begin{align}
\label{definition:qloss:array:gamma}
\chi^{(\kappa,\iota)}_{\gamma}&=\frac{\left[ \left.\ve{F}^{-1}_{\ve{y}}(\ve{\theta})\right|_{K=\kappa} \right]_{11}}{\left[ \left.\ve{\tilde{F}}^{-1}_{\ve{z}}(\ve{\theta}) \right|_{K=\iota} \right]_{11} },\\
\label{definition:qloss:array:zeta}
\chi^{(\kappa,\iota)}_{\zeta}&=\frac{\left[ \left.\ve{F}^{-1}_{\ve{y}}(\ve{\theta}) \right|_{K=\kappa} \right]_{22}}{\left[ \left. \ve{\tilde{F}}^{-1}_{\ve{z}}(\ve{\theta}) \right|_{K=\iota} \right]_{22} }.
\end{align}
Here, the performance of an unquantized system with $\kappa\in\fieldN$ ideal sensors forms the reference for the hard-limiting system with $\iota\in\fieldN$ array elements and is calculated according to \eqref{fisher:matrix:gaussian:full:compact}. These two figures of merit enable analyzing the parameter-specific sensitivity loss when switching from high-resolution A/D conversion at the receivers to low-resolution A/D converters with single-bit output amplitude resolution. 

\pgfplotsset{legend style={rounded corners=2pt,nodes=right}}
\begin{figure}
\centering
\begin{tikzpicture}[scale=1.0]

  	\begin{axis}[ylabel=$\chi^{(K,K)}_{\gamma}$ {[dB]},
  			xlabel=$K$,
			grid,
			ymin=-7,
			ymax=1,
			xmin=2,
			xmax=64,
			legend pos=south east]
																		
			\addplot[red, style=solid, line width=0.75pt,smooth, every mark/.append style={solid}, mark=otimes*, mark repeat=2] table[x index=0, y index=1]{Data/1bitDOA_Sensitivity_vs_K_SNR-6_zeta15.txt};
			\addlegendentry{$\gamma=-\,\,\,6\text{ dB}$}
			
			\addplot[green, style=solid, line width=0.75pt,smooth,every mark/.append style={solid}, mark=square*, mark repeat=2] table[x index=0, y index=1]{Data/1bitDOA_Sensitivity_vs_K_SNR-15_zeta15.txt};
			\addlegendentry{$\gamma=-15\text{ dB}$}
						
			\addplot[blue, style=solid, line width=0.75pt,smooth, every mark/.append style={solid}, mark=diamond*, mark repeat=2] table[x index=0, y index=1]{Data/1bitDOA_Sensitivity_vs_K_SNR-25_zeta15.txt};
			\addlegendentry{$\gamma=-25\text{ dB}$}
												
\end{axis}	
\end{tikzpicture}
\caption{Binary Sensor Array - Hard-limiting Loss vs. Array Size ($\zeta=15^\circ$)}
\label{figure:QLoss:K:gamma}
\end{figure}
\begin{figure}
\centering
\begin{tikzpicture}[scale=1.0]

  	\begin{axis}[ylabel=$\chi^{(K,K)}_{\zeta}$ {[dB]},
  			xlabel=$K$,
			grid,
			ymin=-4.2,
			ymax=-1.8,
			xmin=2,
			xmax=64,
			legend pos=north west]
		
			\addplot[red, style=solid, line width=0.75pt,smooth, every mark/.append style={solid}, mark=otimes*, mark repeat=2] table[x index=0, y index=2]{Data/1bitDOA_Sensitivity_vs_K_SNR-6_zeta15.txt};
			\addlegendentry{$\gamma=-\,\,\,6\text{ dB}$}

			\addplot[green, style=solid, line width=0.75pt,smooth,every mark/.append style={solid}, mark=square*, mark repeat=2] table[x index=0, y index=2]{Data/1bitDOA_Sensitivity_vs_K_SNR-15_zeta15.txt};
			\addlegendentry{$\gamma=-15\text{ dB}$}
						
			\addplot[blue, style=solid, line width=0.75pt,smooth, every mark/.append style={solid}, mark=diamond*, mark repeat=2] table[x index=0, y index=2]{Data/1bitDOA_Sensitivity_vs_K_SNR-25_zeta15.txt};
			\addlegendentry{$\gamma=-25\text{ dB}$}
															
\end{axis}	
\end{tikzpicture}
\caption{Binary Sensor Array - Hard-limiting Loss vs. Array Size ($\zeta=15^\circ$)}
\label{figure:QLoss:K:zeta}
\end{figure}
For three SNR settings and a DOA of $\zeta=15^\circ$, Fig. \ref{figure:QLoss:K:gamma} and Fig. \ref{figure:QLoss:K:zeta} visualize the hard-limiting loss for the two parameters as a function of the array size $K$, while the ideal and the binary system have the same size, i.e., $\kappa=\iota=K$. For both parameters, it can be observed that the hard-limiting loss diminishes when the number of sensors increases while the rate of the improvement depends on the SNR configuration. For example, for the SNR parameter $\gamma$, the coarse quantization loss decreases from $-7.06$ dB ($K=2$) to $-0.66$ dB ($K=64$) when considering an SNR of $-6.0$ dB. Under the same SNR situation, for the DOA parameter $\zeta$, Fig. \ref{figure:QLoss:K:zeta} shows that the loss decreases from $-4.01$ dB ($K=2$) to $-2.54$ dB ($K=64$).

Fig. \ref{figure:gain:K:gamma} and Fig. \ref{figure:gain:K:zeta} show the relative sensitivity gain when comparing to an ideal reference system with high-resolution measurements from $\kappa=2$ sensors. Dashed lines indicate the results when replacing $\ve{z}$ by $\ve{y}$ in \eqref{definition:qloss:array:gamma} and \eqref{definition:qloss:array:zeta}, i.e., comparing an $\infty$-bit system with $K$ sensors to an ideal reference with $\kappa=2$ sensors. It can be seen that the performance of the binary sensor array scales with the array size $K$ in the same way than for the unquantized sensor system. The results also show that a binary sensor array with $K$ elements can outperform an ideal system with an array size of $\frac{K}{2}$. This insight corroborates that, in low and medium SNR regimes, array processing with hardware and energy-efficient binary sensor architectures can be performed at high accuracy if the number of array elements is sufficiently large. Note that here the Fisher information lower bound \eqref{quantitative:fisher:bound} allows to mathematically explore the Fisher matrix $\ve{F_z}(\ve{\theta})$ of the $1$-bit system for cases with $K>2$, while in \cite{BarShalom02}, the mathematical analysis had to be restricted to $K=2$ due to the missing expression for the orthant probabilities with $M>4$.

To analyze the quality of the heuristic Gaussian approximation \eqref{approximate:fisher:hardlimited} for the conservative information matrix \eqref{pessimistic:fisher:information:matrix}, in Fig. \ref{figure:QLoss:SNR} the quantization loss with $\kappa=\iota=32$ is depicted as a function of the SNR. It can be seen that the approximation provides high-quality results for SNR values below $-15$ dB. Above this point, the approximation \eqref{approximate:fisher:hardlimited} here leads to results that overestimate the inference capability with the binary data model. When considering the estimation error for the SNR parameter $\gamma$ at $-5$ dB, the approximated conservative information matrix \eqref{approximate:fisher:hardlimited} indicates that the binary system outperforms the ideal sensor array, which violates the fundamental information-theoretic principle \eqref{fisher:inequality:data:processing}. This indicates that, for signal parameter estimation problems with binary measurements, approximations that take advantage of the favorable structure of the Gaussian distribution must be considered with caution. 
\begin{figure}
\centering
\begin{tikzpicture}[scale=1.0]

  	\begin{axis}[ylabel=$\chi^{(2,K)}_{\gamma}$ {[dB]},
  			xlabel=$K$,
			grid,
			ymin=-9,
			ymax=25,
			xmin=2,
			xmax=32,
			legend pos=south east]
			
			\addplot[blue, style=dashed, line width=0.75pt, smooth, every mark/.append style={solid}, mark=diamond, mark repeat=2] table[x index=0, y index=3]{Data/1bitDOA_Sensitivity_vs_K_SNR-25_zeta15.txt};
			\addlegendentry{$\gamma=-25\text{ dB}$ ($\infty$-bit)}
			
			\addplot[blue, style=solid, line width=0.75pt,smooth, every mark/.append style={solid}, mark=diamond*, mark repeat=2] table[x index=0, y index=5]{Data/1bitDOA_Sensitivity_vs_K_SNR-25_zeta15.txt};
			\addlegendentry{$\gamma=-25\text{ dB}$}
			
			\addplot[red, style=dashed, line width=0.75pt,smooth, every mark/.append style={solid}, mark=o, mark repeat=2] table[x index=0, y index=3]{Data/1bitDOA_Sensitivity_vs_K_SNR-6_zeta15.txt};
			\addlegendentry{$\gamma=-\,\,\,6\text{ dB}$ ($\infty$-bit)}
														
			\addplot[red, style=solid, line width=0.75pt,smooth, every mark/.append style={solid}, mark=otimes*, mark repeat=2] table[x index=0, y index=5]{Data/1bitDOA_Sensitivity_vs_K_SNR-6_zeta15.txt};
			\addlegendentry{$\gamma=-\,\,\,6\text{ dB}$}
												
\end{axis}	
\end{tikzpicture}
\caption{Binary Sensor Array - Performance Gain vs. Array Size ($\zeta=15^\circ$)}
\label{figure:gain:K:gamma}
\end{figure}
\begin{figure}
\centering
\begin{tikzpicture}[scale=1.0]

  	\begin{axis}[ylabel=$\chi^{(2,K)}_{\zeta}$ {[dB]},
  			xlabel=$K$,
			grid,
			ymin=-5,
			ymax=55,
			xmin=2,
			xmax=32,
			legend pos=south east]
			
			\addplot[blue, style=dashed, line width=0.75pt, smooth, every mark/.append style={solid}, mark=diamond, mark repeat=2] table[x index=0, y index=4]{Data/1bitDOA_Sensitivity_vs_K_SNR-25_zeta15.txt};
			\addlegendentry{$\gamma=-25\text{ dB}$ ($\infty$-bit)}
			
			\addplot[blue, style=solid, line width=0.75pt, smooth, every mark/.append style={solid}, mark=diamond*, mark repeat=2] table[x index=0, y index=6]{Data/1bitDOA_Sensitivity_vs_K_SNR-25_zeta15.txt};
			\addlegendentry{$\gamma=-25\text{ dB}$}
				
			\addplot[red, style=dashed, line width=0.75pt, smooth, every mark/.append style={solid}, mark=o, mark repeat=2] table[x index=0, y index=4]{Data/1bitDOA_Sensitivity_vs_K_SNR-6_zeta15.txt};
			\addlegendentry{$\gamma=-\,\,\,6\text{ dB}$ ($\infty$-bit)}
							
			\addplot[red, style=solid, line width=0.75pt, smooth, every mark/.append style={solid}, mark=otimes*, mark repeat=2] table[x index=0, y index=6]{Data/1bitDOA_Sensitivity_vs_K_SNR-6_zeta15.txt};
			\addlegendentry{$\gamma=-\,\,\,6\text{ dB}$}
							
\end{axis}	
\end{tikzpicture}
\caption{Binary Sensor Array - Performance Gain vs. Array Size ($\zeta=15^\circ$)}
\label{figure:gain:K:zeta}
\end{figure}
\begin{figure}[!htbp]
\centering
\begin{tikzpicture}[scale=1.0]

  	\begin{axis}[ylabel=$\chi$ {[dB]},
  			xlabel=$\gamma$ {[dB]},
			grid,
			ymin=-5,
			ymax=1,
			xmin=-40,
			xmax=10,
			legend pos=north west]
			
			\addplot[red, style=dashed, line width=0.75pt, smooth, every mark/.append style={solid}, mark=o, mark repeat=4, mark phase=3] table[x index=0, y index=1]{Data/1bitDOA_ApproximateSensitivity_vs_SNR_K32_zeta15.txt};
			\addlegendentry{$\chi^{(32,32)}_\gamma$ (approx.)}
																		
			\addplot[red, style=solid, line width=0.75pt, smooth, every mark/.append style={solid}, mark=otimes*, mark repeat=4] table[x index=0, y index=1]{Data/1bitDOA_Sensitivity_vs_SNR_K32_zeta15.txt};
			\addlegendentry{$\chi^{(32,32)}_\gamma$}
			
			\addplot[blue, style=dashed, line width=0.75pt, smooth,every mark/.append style={solid}, mark=square, mark repeat=4, mark phase=3] table[x index=0, y index=2]{Data/1bitDOA_ApproximateSensitivity_vs_SNR_K32_zeta15.txt};
			\addlegendentry{$\chi^{(32,32)}_\zeta$ (approx.)}
			
			\addplot[blue, style=solid, line width=0.75pt, smooth,every mark/.append style={solid}, mark=square*, mark repeat=4] table[x index=0, y index=2]{Data/1bitDOA_Sensitivity_vs_SNR_K32_zeta15.txt};
			\addlegendentry{$\chi^{(32,32)}_\zeta$}
											
\end{axis}	
\end{tikzpicture}
\caption{Binary Sensor Array - Hard-limiting Loss vs. SNR ($\zeta=15^\circ$)}
\label{figure:QLoss:SNR}
\end{figure}

However, here such an approximation provides some mathematical insight into the information loss in the low SNR regime, where the correlation matrix of the unquantized data approaches identity, i.e.,
\begin{align}
\ve{C}_{\ve{y}}(\ve{\theta})\to\ve{I}.
\end{align}
For the covariance matrix of the binary data one then obtains
\begin{align}\label{covariance:quantized:low:snr}
\ve{R}_{\ve{z}}(\ve{\theta})\approx \frac{2}{\pi} \arcsin{\ve{I}}=\ve{I}
\end{align}
and for its derivatives
\begin{align}
&\left[ \frac{\partial \ve{R}_{\ve{z}}(\ve{\theta})}{\partial \theta_d} \right]_{ij} \approx  \frac{2}{\pi} \left[ \frac{\partial \ve{C}_{\ve{y}}(\ve{\theta})}{\partial \theta_d} \right]_{ij}, \quad i \neq j.
\end{align}
Therefore, the conservative information matrix for the binary sensor system and the Fisher information matrix of the ideal $\infty$-bit system can be related via the approximation
\begin{align}\label{approximate:fisher:hardlimited:low:snr}
\ve{\tilde{F}_z}(\ve{\theta}) &\approx \frac{1}{2}  \bigg( \frac{\partial \ve{r}_{\ve{z}}(\ve{\theta})}{\partial \ve{\theta}} \bigg)^{\T}  \big(\ve{R}_{\ve{z}}^{-1}(\ve{\theta}) \otimes \ve{R}_{\ve{z}}^{-1}(\ve{\theta}) \big) \frac{\partial \ve{r}_{\ve{z}}(\ve{\theta})}{\partial \ve{\theta}}\\
&\approx \frac{1}{2} \left(\frac{2}{\pi}\right)^2 \bigg( \frac{\partial \ve{r}_{\ve{y}}(\ve{\theta})}{\partial \ve{\theta}} \bigg)^{\T}  \frac{\partial \ve{r}_{\ve{y}}(\ve{\theta})}{\partial \ve{\theta}}\\
&\approx \left(\frac{2}{\pi}\right)^2 \ve{F_y}(\ve{\theta}).
\end{align}
This shows that for shift in variance problems like the DOA estimation task considered here, the hard-limiting loss with low SNR is approximately $10\log\left(\frac{2}{\pi}\right)^2 = -3.92$ dB. Note that this stands in contrast to shift in mean problems where the low SNR $1$-bit loss is known to be $10\log\left(\frac{2}{\pi}\right) = -1.96$ dB.  
\subsection{Statistical Inference with Recursive Binary Sampling}
As a second engineering example, the task of signal parameter estimation using digital measurements obtained by a recursive binary sampling scheme is considered. In contrast to hard-limiting \eqref{mapping:sign}, the quantization error of each sample is fed back and thus taken into account in subsequent sampling values. Such a signal acquisition technique is found in A/D converters employing $\Sigma\Delta$-modulation. In comparison to flash architectures, the advantage of such A/D converters is that high-resolution signal digitization can be realized using one comparator and a feedback loop. To this end, the high-resolution samples of the input signal are reconstructed by digital decimation and filtering of the binary measurement data, which requires significant oversampling (typically by a factor of $16$ to $256$) to achieve a sufficiently low signal distortion level. In several sensing applications, however, the task of the digital processing unit is to infer a few signal parameters. In such cases, reconstruction of the high-resolution samples of the analog input to the A/D converter is not necessarily required. With a probabilistic system model $p_{\ve{z}}(\ve{z};\ve{\theta})$ of the digitization process at hand, the signal parameters can be extracted directly from the binary measurement data through an appropriate statistical processing method. 

In the following, the parameter-specific information loss caused by recursive binary sampling when following such a direct digital processing approach is assessed. Since probabilistic modeling of the binary measurements is challenging, a data-driven method is employed. By simulating the recursive signal acquisition process with appropriate random input data, the mean \eqref{equivalence:mean} and the covariance matrix \eqref{equivalence:covariance} of user-defined statistics at the output are obtained. Such a procedure allows estimating a conservative approximation for the Fisher information matrix \eqref{pessimistic:fisher:information:matrix} and, therefore, assessing the guaranteed achievable estimation performance. In the context of a single-sensor parameter estimation problem, the obtained results are used to identify favorable feedback weights and oversampling factors.

For the data-driven analysis, the analog sensor signal
\begin{align}\label{anaolg:signal:model:sigma:delta}
y(t)=\mu+\eta(t)
\end{align}
is considered, where $\mu\in\fieldR$ is a constant offset signal and $\eta(t)\in\fieldR$ a band-limited zero-mean Gaussian process with autocorrelation
\begin{align}
r_{\eta}(t)=\sigma^2 \sincb{\frac{\Omega_{\eta}}{\pi} t},
\end{align}
for which $\Omega_{\eta}\in\fieldR, \Omega_{\eta}>0,$ denotes the spectral bandwidth in radians per second. For A/D conversion, it is assumed that an observation window of duration $T_0$ is available within which the analog sensor signal \eqref{anaolg:signal:model:sigma:delta} is sampled $M$ times in equidistant intervals $T_s$. The rate of the sampling process $\Omega_s=\frac{2\pi}{T_s}$ in relation to the bandwidth $\Omega_{\eta}$ of the noise process is
\begin{align}
\Omega_s=2\Omega_\eta \lambda,
\end{align}
with $\lambda\in\fieldR,\lambda \geq 1,$ being the oversampling factor. Fixing the total observation time $T_0$ results in $M=\floor{\lambda M_0}$, where $M_0$ is the number of samples obtained without oversampling (i.e., $\lambda=1$). For simplicity, the analog signal bandwidth is normalized to $\Omega_\eta=\pi$, such that $T_s=\frac{1}{\lambda}$. Assuming an ideal A/D converter with infinite amplitude resolution, results in samples
\begin{align}
y_m=y(T_s m)=y\Big(\frac{m}{ \lambda} \Big),
\end{align}
such that the digital measurement data
\begin{align}
\ve{y}=\begin{bmatrix} y_1 &\ldots & y_M \end{bmatrix}^{\T}
\end{align}
follows the multivariate Gaussian distribution model
\begin{align}\label{digital:signal:model:sigma:delta:highres}
\ve{y} \sim\distN(\mu\ve{1},\sigma^2 \ve{C}_{\ve{\eta}}),
\end{align}
where the entries of the noise correlation matrix are
\begin{align}
\left[ \ve{C}_{\ve{\eta}} \right]_{ij} = \sincb{\frac{|i-j|}{\lambda}},
\end{align}
while $\ve{1}$ denotes the all-ones vector.

With $\Sigma\Delta$-modulation featuring a first-order error feedback loop, the sampling process (see Fig. \ref{adc:sigma:delta}) can be described by \cite{Daubechies03}
\begin{align}
\label{sigma:delta:scheme:first:order:internal}
s_m&=y\Big(\frac{m}{ \lambda} \Big)+\alpha s_{m-1}-z_m,\\
\label{sigma:delta:scheme:first:order:external}
z_m&=\sign{y\Big(\frac{m}{ \lambda} \Big)+\alpha s_{m-1}},
\end{align}
where $s_m$ constitutes an internal variable of the converter which is initialized $s_0=0$. 

\begin{figure}[!htbp]
\centering

	\begin{tikzpicture}[scale=0.8]

		\draw [thick,rounded corners=2pt,->] (1,1) -- (2,1);
		\draw (1,1) node[left] {$y(\frac{m}{\lambda})$};

 		\draw (2.25,1) circle (0.25cm);
		\draw[color=black,line width=0.5] (2.25,1+0.25) -- (2.25,1-0.25);
  		\draw[color=black,line width=0.5] (2.25+0.25,1) -- (2.25-0.25,1);
		
		\draw [thick,rounded corners=2pt,->] (2.5,1) -- (4.5,1);
		
		\draw [thick,rounded corners=2pt] (4.5,0.25) rectangle (7.5,1.75);
		\draw [thick] (5,0.5) -- (6,0.5) --  (6,1.5)  --  (7,1.5);
		\draw[dotted] (5,1) -- (7,1);
		\draw[dotted] (6,1.6) -- (6,0.4);
		
		\draw [thick,rounded corners=2pt,->] (7.5,1) -- (9.5,1);
		\draw (9.5,1) node[right] {$z_m$};
		
		\draw [thick,rounded corners=2pt,->] (3.5,1) -- (3.5,-2.5);
		\draw[fill] (3.5,1) circle (0.075cm);
		
 		\draw (3.5,-1.25 - 1.5) circle (0.25cm);
		\draw[color=black,line width=0.5] (3.5,-1.25+0.25 - 1.5) -- (3.5,-1.25-0.25 - 1.5);
  		\draw[color=black,line width=0.5] (3.5+0.25,-1.25 - 1.5) -- (3.5-0.25,-1.25 - 1.5);
		
		\draw [thick,rounded corners=2pt,->] (8.5,1) -- (8.5,-1.25 - 1.5) -- (3.75,-1.25 - 1.5);
		\draw[fill] (8.5,1) circle (0.075cm);
		\draw (3.75,-1.25 - 1.5) node[below right] {$-$};
		
		\draw [thick,rounded corners=2pt] (1.75,0.25) rectangle (2.75,-0.75);
		\draw (2.25,-0.25) node {\scriptsize{Gain $\alpha$}};
		
		\draw [thick,rounded corners=2pt] (1.75,0.25 - 1.5) rectangle (2.75,-0.75 - 1.5);
		\draw (2.25,-0.25 - 1.5) node {\scriptsize{Delay}};

		\draw [thick,rounded corners=2pt,->] (3.5-0.25,-1.25 - 1.5) -- (2.25,-1.25 - 1.5) -- (2.25,-0.75 - 1.5);
		\draw (2.25,-1.25 - 1.5) node[below] {$s_m$};
		
		\draw [thick,rounded corners=2pt,->] (2.25,0.25) -- (2.25,0.75);
		\draw [thick,rounded corners=2pt,->] (2.25,0.25-1.5) -- (2.25,0.75-1.5);

	\end{tikzpicture}
	\caption{Analog-to-Digital Converter with $\Sigma\Delta$-Modulation}
	\label{adc:sigma:delta}
\end{figure}

The error feedback weight $\alpha\in\fieldR$ is open to design, while setting $\alpha=0$ results in the low-complexity $1$-bit sampling scheme considered in Section \ref{section:binary:array:processing}, i.e., 
\begin{align}\label{sigma:delta:scheme:first:order:zero:feedback}
z_m&=\sign{y\Big(\frac{m}{ \lambda} \Big)}.
\end{align}
Probabilistic modeling of the binary measurements acquired with the recursive digitization process \eqref{sigma:delta:scheme:first:order:external} turns out to be challenging. On the one hand, like with hard-limiting \eqref{sigma:delta:scheme:first:order:zero:feedback}, the digital data follows a multivariate binary distribution
\begin{align}\label{binary:distribution:sigma:delta}
\ve{z} \sim p_{\ve{z}}(\ve{z};\ve{\theta}),
\end{align}
modulated by the analog signal parameters 
\begin{align}\label{parameters:sigma:delta}
\ve{\theta}=\begin{bmatrix} \mu &\sigma^2 \end{bmatrix}^{\T},
\end{align}
and as such is intractable. Avoiding the explicit formulation of the likelihood \eqref{binary:distribution:sigma:delta} by using an equivalent exponential family and mathematically determining the mean \eqref{equivalence:mean} and the covariance matrix \eqref{equivalence:covariance} of certain statistics \eqref{definition:auxiliary:statistics} is also demanding as the error feedback loop of the converter \eqref{sigma:delta:scheme:first:order:internal} introduces dependencies between the samples which are difficult to take into account.

Therefore, calibrated binary measurements
\begin{align}\label{dataset:learning}
\ve{Z}=\begin{bmatrix} \ve{z}_1 &\ve{z}_2 &\ldots &\ve{z}_N \end{bmatrix}
\end{align}
are generated by Monte-Carlo simulations of the recursive data acquisition process \eqref{sigma:delta:scheme:first:order:external} with input samples from the unquantized distribution model $p_{\ve{y}}(\ve{y};\ve{\theta})$. Using the sufficient statistics of the quadratic exponential family
\begin{align}
\ve{\tilde{\phi}}(\ve{z})=\begin{bmatrix} \ve{z}\\ \ve{\Phi}\vec{\ve{z} \ve{z}^{\T}}\end{bmatrix},
\end{align}
the empirical mean
\begin{align}
\ve{\hat{\mu}}_{\ve{\tilde{\phi}}}(\ve{Z}|\ve{\theta}) = \frac{1}{N} \sum_{n=1}^{N} \ve{\tilde{\phi}} (\ve{z}_n)
\end{align}
and the empirical covariance matrix
\begin{align}
\ve{\hat{R}}_{\ve{\tilde{\phi}}}(\ve{Z}|\ve{\theta}) = \frac{1}{N} \sum_{n=1}^{N} \ve{\tilde{\phi}} (\ve{z}_n) \ve{\tilde{\phi}}^{\T} (\ve{z}_n) - \ve{\hat{\mu}}_{\ve{\tilde{\phi}}}(\ve{Z}|\ve{\theta})\ve{\hat{\mu}}^{\T}_{\ve{\tilde{\phi}}}(\ve{Z}|\ve{\theta})
\end{align}
are computed. With the symmetric finite difference
\begin{align}
\frac{\partial \ve{\hat{\mu}}_{\ve{\tilde{\phi}}}(\ve{Z}|\ve{\theta})}{ \partial \theta_d} \approx \frac{ \ve{\hat{\mu}}_{\ve{\tilde{\phi}}}(\ve{Z}|\ve{\theta}+\ve{\Delta}^{(d)}) -\ve{\hat{\mu}}_{\ve{\tilde{\phi}}}(\ve{Z}|\ve{\theta}-\ve{\Delta}^{(d)}) }{2\Delta},
\end{align}
where $\ve{\Delta}^{(d)}$ denotes an all-zero vector with the $d$th element set to $\Delta\in\fieldR, \Delta>0$, the conservative information matrix \eqref{pessimistic:fisher:information:matrix} is then approximated empirically by
\begin{align}\label{approximation:fisher:matrix:data:driven}
\ve{\tilde{F}_z}(\ve{\theta}) & \approx \bigg(\frac{\partial \ve{\hat{\mu}}_{\ve{\tilde{\phi}}}(\ve{Z}|\ve{\theta})}{ \partial \ve{\theta}} \bigg)^{\T} \ve{\hat{R}}_{\ve{\tilde{\phi}}}^{-1}(\ve{Z}|\ve{\theta}) \frac{\partial \ve{\hat{\mu}}_{\ve{\tilde{\phi}}}(\ve{Z}|\ve{\theta})}{ \partial \ve{\theta}}\notag\\
&=\ve{\hat{F}_z}(\ve{\theta}).
\end{align}
Note that this data-driven approximation $\ve{\hat{F}_z}(\ve{\theta})$ of the conservative information matrix $\ve{\tilde{F}_z}(\ve{\theta})$ requires $2D+1$ calibrated datasets \eqref{dataset:learning} and an appropriate choice of the finite difference parameter $\Delta$. For all following cases, $M_0=10$,  $N=10^8$, and $\Delta=0.01$ are used. 

For the assessment of the recursive binary sampling loss under the signal model \eqref{anaolg:signal:model:sigma:delta} the empirical loss measures
\begin{align}\label{definition:qloss:sigma:delta}
\hat{\chi}_{\mu}&=\frac{\left[ \ve{F}^{-1}_{\ve{y}}(\ve{\theta}) \right]_{11}}{\left[ \ve{\hat{F}}^{-1}_{\ve{z}}(\ve{\theta}) \right]_{11} },\\
\hat{\chi}_{\sigma^2}&=\frac{\left[ \ve{F}^{-1}_{\ve{y}}(\ve{\theta}) \right]_{22}}{\left[ \ve{\hat{F}}^{-1}_{\ve{z}}(\ve{\theta}) \right]_{22} },
\end{align}
are defined. With the notational conventions
\begin{align}
\ve{c}_{\ve{\eta}}(\ve{\theta})=\vec{\ve{C}_{\ve{\eta}}(\ve{\theta})},
\end{align}
and
\begin{align}
\ve{R}_{\ve{\eta}}(\ve{\theta})=\sigma^2\ve{C}_{\ve{\eta}}(\ve{\theta}),
\end{align}
the Fisher information matrix of the unquantized input data is calculated
\begin{align}
\ve{F}_{\ve{y}}(\ve{\theta})=\begin{bmatrix} \ve{1} &\ve{0}\\ \ve{0} &\ve{c}_{\ve{\eta}} \end{bmatrix}^{\T}  \begin{bmatrix} \ve{R}^{-1}_{\ve{\eta}} &\ve{0}\\
\ve{0} & \frac{1}{2} (\ve{R}^{-1}_{\ve{\eta}} \otimes \ve{R}^{-1}_{\ve{\eta}} ) \end{bmatrix} \begin{bmatrix} \ve{1} &\ve{0}\\ \ve{0} &\ve{c}_{\ve{\eta}} \end{bmatrix}.
\end{align}

Fig. \ref{fig:feedback:opt:input:mean:zero:mean} and Fig. \ref{fig:feedback:opt:input:variance:zero:mean} show the empirical information loss of the recursive binary sampling process \eqref{sigma:delta:scheme:first:order:external} as a function of the error feedback weight $\alpha$ in a situation where the signal mean is $\mu=0$, and the noise variance is $\sigma^2=1$. For the mean parameter $\mu$, it is observed that error feedback optimization enables significantly reducing the binary sampling loss while oversampling provides minor performance improvements. For the variance $\sigma^2$, the sensitivity loss is more pronounced than for the mean parameter $\mu$. However, oversampling together with an adjusted design of the feedback weight results in a significantly higher parameter-specific sensitivity. Under an oversampling factor $\lambda=4$, for the mean the favorable error feedback weight is $\alpha\approx 1$, while for the noise variance, $\alpha\approx 1.05$ leads to the best performance.
\begin{figure}
\centering
\begin{tikzpicture}[scale=1.0]

  	\begin{axis}[ylabel=$\hat{\chi}_{\mu}$ {[dB]},
  			xlabel=$\text{Error Feedback Weight }\alpha$,
			grid,
			ymin=-5.0,
			ymax=1.0,
			xmin=0.4,
			xmax=1.6,
			legend pos=south west]
						
			\addplot[red, style=solid, line width=0.75pt,smooth, every mark/.append style={solid}, mark=otimes*, mark repeat=1] table[x index=0, y index=1]{Data/SigmaDelta_Sensitivity_vs_Feedback_lambda1_mean0_variance1_delta0.01.txt};
			\addlegendentry{$\lambda=1$}

			\addplot[blue, style=solid, line width=0.75pt,smooth,every mark/.append style={solid}, mark=square*, mark repeat=1] table[x index=0, y index=1]{Data/SigmaDelta_Sensitivity_vs_Feedback_lambda2_mean0_variance1_delta0.01.txt};
			\addlegendentry{$\lambda=2$}
									
			\addplot[green, style=solid, line width=0.75pt,smooth, every mark/.append style={solid}, mark=diamond*, mark repeat=1] table[x index=0, y index=1]{Data/SigmaDelta_Sensitivity_vs_Feedback_lambda4_mean0_variance1_delta0.01.txt};
			\addlegendentry{$\lambda=4$}
											
\end{axis}	
\end{tikzpicture}
\caption{$\Sigma\Delta$-Modulation - Sensitivity vs. Feedback Design ($\mu=0, \sigma^2=1$)}
\label{fig:feedback:opt:input:mean:zero:mean}
\end{figure}
\begin{figure}
\centering
\begin{tikzpicture}[scale=1.0]

  	\begin{axis}[ylabel=$\hat{\chi}_{\sigma^2}$ {[dB]},
  			xlabel=$\text{Error Feedback Weight }\alpha$,
			grid,
			ymin=-16.0,
			ymax=1.0,
			xmin=0.4,
			xmax=1.6,
			legend pos=south east]
						
			\addplot[red, style=solid, line width=0.75pt,smooth, every mark/.append style={solid}, mark=otimes*, mark repeat=1] table[x index=0, y index=2]{Data/SigmaDelta_Sensitivity_vs_Feedback_lambda1_mean0_variance1_delta0.01.txt};
			\addlegendentry{$\lambda=1$}

			\addplot[blue, style=solid, line width=0.75pt,smooth,every mark/.append style={solid}, mark=square*, mark repeat=1] table[x index=0, y index=2]{Data/SigmaDelta_Sensitivity_vs_Feedback_lambda2_mean0_variance1_delta0.01.txt};
			\addlegendentry{$\lambda=2$}
									
			\addplot[green, style=solid, line width=0.75pt,smooth, every mark/.append style={solid}, mark=diamond*, mark repeat=1] table[x index=0, y index=2]{Data/SigmaDelta_Sensitivity_vs_Feedback_lambda4_mean0_variance1_delta0.01.txt};
			\addlegendentry{$\lambda=4$}

\end{axis}	
\end{tikzpicture}
\caption{$\Sigma\Delta$-Modulation - Sensitivity vs. Feedback Design ($\mu=0, \sigma^2=1$)}
\label{fig:feedback:opt:input:variance:zero:mean}
\end{figure}

Fig. \ref{fig:feedback:opt:input:mean:unit:mean} and Fig. \ref{fig:feedback:opt:input:variance:unit:mean} depict the results for $\mu=0.5$ and $\sigma^2=1$. For the mean parameter, the binary sampling loss with error feedback weights $\alpha<0.8$ is more pronounced than for the situation with $\mu=0$, while the optimal feedback weight for $\lambda=4$ shifts towards $\alpha\approx 0.95$. For the noise variance, the achievable sensitivity is similar to the case with $\mu=0$ and significantly improves through oversampling, while the optimal error feedback design for $\lambda=4$ is $\alpha\approx 0.95$.

In both considered cases, the empirical results show that high-resolution signal parameter estimation from $\Sigma\Delta$-measurements with $1$-bit amplitude resolution is possible if moderate oversampling (e.g., $\lambda=4$) and optimized error feedback weights are used. Note, however, that in contrast to the standard approach where deterministic reconstruction of the high-resolution samples is performed before solving the statistical inference task in the conventional Gaussian framework \eqref{digital:signal:model:sigma:delta:highres}, a direct processing technique requires probabilistic modeling of the binary sensor measurements and iteratively solving the likelihood-oriented optimization problem \eqref{definition:cmle}.
\begin{figure}
\centering
\begin{tikzpicture}[scale=1.0]

  	\begin{axis}[ylabel=$\hat{\chi}_{\mu}$ {[dB]},
  			xlabel=$\text{Error Feedback Weight }\alpha$,
			grid,
			ymin=-5.0,
			ymax=1.0,
			xmin=0.4,
			xmax=1.6,
			legend pos=north east]
						
			\addplot[red, style=solid, line width=0.75pt,smooth, every mark/.append style={solid}, mark=otimes*, mark repeat=1] table[x index=0, y index=1]{Data/SigmaDelta_Sensitivity_vs_Feedback_lambda1_mean0.5_variance1_delta0.01.txt};
			\addlegendentry{$\lambda=1$}

			\addplot[blue, style=solid, line width=0.75pt,smooth,every mark/.append style={solid}, mark=square*, mark repeat=1] table[x index=0, y index=1]{Data/SigmaDelta_Sensitivity_vs_Feedback_lambda2_mean0.5_variance1_delta0.01.txt};
			\addlegendentry{$\lambda=2$}
									
			\addplot[green, style=solid, line width=0.75pt,smooth, every mark/.append style={solid}, mark=diamond*, mark repeat=1] table[x index=0, y index=1]{Data/SigmaDelta_Sensitivity_vs_Feedback_lambda4_mean0.5_variance1_delta0.01.txt};
			\addlegendentry{$\lambda=4$}

\end{axis}	
\end{tikzpicture}
\caption{$\Sigma\Delta$-Modulation - Sensitivity vs. Feedback Design ($\mu=0.5, \sigma^2=1$)}
\label{fig:feedback:opt:input:mean:unit:mean}
\end{figure}
\begin{figure}
\centering
\begin{tikzpicture}[scale=1.0]

  	\begin{axis}[ylabel=$\hat{\chi}_{\sigma^2}$ {[dB]},
  			xlabel=$\text{Error Feedback Weight }\alpha$,
			grid,
			ymin=-16.0,
			ymax=1.0,
			xmin=0.4,
			xmax=1.6,
			legend pos=south east]
						
			\addplot[red, style=solid, line width=0.75pt,smooth, every mark/.append style={solid}, mark=otimes*, mark repeat=1] table[x index=0, y index=2]{Data/SigmaDelta_Sensitivity_vs_Feedback_lambda1_mean0.5_variance1_delta0.01.txt};
			\addlegendentry{$\lambda=1$}

			\addplot[blue, style=solid, line width=0.75pt,smooth,every mark/.append style={solid}, mark=square*, mark repeat=1] table[x index=0, y index=2]{Data/SigmaDelta_Sensitivity_vs_Feedback_lambda2_mean0.5_variance1_delta0.01.txt};
			\addlegendentry{$\lambda=2$}
				
			\addplot[green, style=solid, line width=0.75pt,smooth, every mark/.append style={solid}, mark=diamond*, mark repeat=1] table[x index=0, y index=2]{Data/SigmaDelta_Sensitivity_vs_Feedback_lambda4_mean0.5_variance1_delta0.01.txt};
			\addlegendentry{$\lambda=4$}

\end{axis}	
\end{tikzpicture}
\caption{$\Sigma\Delta$-Modulation - Sensitivity vs. Feedback Design ($\mu=0.5, \sigma^2=1$)}
\label{fig:feedback:opt:input:variance:unit:mean}
\end{figure}
\section{Conclusion}
An approach for the derivation of compact and tractable lower bounds for the Fisher information matrix of arbitrary random data models $p_{\ve{u}}(\ve{u};\ve{\theta})$ has been discussed. To this end, the system model is approximated by an exponential family distribution $\tilde{p}_{\ve{u}}(\ve{u};\ve{\theta})$ with sufficient statistics $\ve{\tilde{\phi}}(\ve{u})$. The statistics feature equivalent mean and covariance under both probabilistic models. Such an surrogate model exhibits lower Fisher information (in a matrix sense) than the true data-generating model and, therefore, yields a conservative approximation $\ve{\tilde{F}_u}(\ve{\theta})$ for the Fisher information matrix $\ve{F_u}(\ve{\theta})$. For the computation of the conservative information matrix $\ve{\tilde{F}_u}(\ve{\theta})$, the involved probabilistic models $p_{\ve{u}}(\ve{u};\ve{\theta})$ and $\tilde{p}_{\ve{u}}(\ve{u};\ve{\theta})$ do not have to be explicitly characterized. Required are solely the mean and covariance matrix of certain statistics $\ve{\tilde{\phi}}(\ve{u})$, which can be obtained mathematically or estimated from calibrated data. Through the free choice of the statistics $\ve{\tilde{\phi}}(\ve{u})$, the user has control over the complexity of these tasks. Further, mean and covariance allow evaluating the score function of the surrogate model $\tilde{p}_{\ve{u}}(\ve{u};\ve{\theta})$, such that a consistent estimator, asymptotically achieving the error level defined by the inverse of the conservative information matrix $\ve{\tilde{F}_u}(\ve{\theta})$, can be designed without having access to a likelihood function. Considering two electrical engineering applications requiring sensor system design specification for signal parameter estimation with binary measurements, it was demonstrated how to exploit the discussed Fisher information approximation framework for a quantitative hardware-aware and parameter-specific sensitivity analysis. The presented results corroborate that low-complexity sampling schemes preserve the capability to conduct parameter estimation with the acquired binary data at high performance and, therefore, have the potential to constitute the key technology for modern and advanced sensing architectures.
\section{Acknowledgement}
Useful discussions concerning the properties of exponential family distributions with Kurt Barb\'{e} (Vrije Universiteit Brussel) as well as fruitful discussions on hardware-aware signal processing and low-complexity $1$-bit A/D conversion with Josef A. Nossek (Technische Universit\"at M\"unchen) are gratefully acknowledged. Their comments on the manuscript have helped to improve the presentation.
\bibliographystyle{IEEEbib}

\end{document}